%% file: main.tex
\newif\ifanonymized
\anonymizedfalse    

\newif\ifsubmit
\submitfalse	
\submittrue	

\documentclass[nonacm]{acmart}

\input{preamble}
    \usepackage{algpseudocode}
    \usepackage{amsmath}
    \usepackage{subcaption}
    \usepackage{tablefootnote}
    \usepackage{enumitem}

\begin{document}

\title[Imputation Matters: A Deeper Look into an Overlooked Step in Longitudinal Health and Behavior  Sensing]{Imputation Matters: A Deeper Look into an Overlooked Step in Longitudinal Health and Behavior Sensing Research}
\date{}

\ifanonymized
   \author{Anonymized for blind submission}
\else

\author{Akshat Choube}
        \email{choube.a@northeastern.edu}
        \affiliation{%
          \institution{Northeastern University}
          \country{United States}
        }
        
        \author{Rahul Majethia}
        \email{rmajethia@acm.org}
        \affiliation{%
          \institution{Association for Computing Machinery}
          \country{India}
        }

          \author{Sohini Bhattacharya}
        \email{sb355@snu.edu.in}
        \affiliation{%
          \institution{Shiv Nadar Institution of Eminence}
          \country{India}
        }

         \author{Vedant Das Swain}
     \email{v.dasswain@northeastern.edu}
        \affiliation{%
          \institution{Northeastern University}
          \country{United States}
        }

         \author{Jiachen Li}
     \email{li.jiachen4@northeastern.edu}
        \affiliation{%
          \institution{Northeastern University}
          \country{United States}
        }
     
     \author{Varun Mishra}
     \email{v.mishra@northeastern.edu}
        \affiliation{%
          \institution{Northeastern University}
          \country{United States}
        }
        
\fi



\settopmatter{printacmref=false}


\input{abstract}		
\maketitle  

\input{headings}	
\input{body}		
\input{ack}			


\bibliographystyle{ACM-Reference-Format}

\bibliography{bibs/local}

\end{document}

%% file: preamble.tex
\usepackage{fancyhdr}

\usepackage{url}

\usepackage{color}
\usepackage[ruled,vlined]{algorithm2e}
\usepackage{colortbl}

\usepackage[mmddyyyy,hhmmss]{datetime}

\usepackage{xspace}

\usepackage{graphicx}
\DeclareGraphicsExtensions{.pdf,.png,.eps,.ps,.jpg}

\ifsubmit
  \newcommand{\hey}[1]{\relax}
  \newcommand{\heyvarun}[1]{\relax}
  \newcommand{\heyakshat}[1]{\relax}
  \newcommand{\bibtex}[1]{\relax}
  \newcommand{\note}[1]{\relax}
\else
  \newcommand{\hey}[1]{\textcolor{magenta}{[{#1}]}}
  \newcommand{\heyvarun}[1]{\textcolor{blue}{[Varun: {#1}]}}
  \newcommand{\heyakshat}[1]{\textcolor{orange}{[Akshat: {#1}]}}
  \newcommand{\note}[1]{\par\textcolor{magenta}{Note: {#1}}\par}
  \newcommand{\bibtex}[1]{\textcolor{red}{@bibtex}\{#1\}} 
\fi

\newcommand{\hide}[1]{\relax}

\newcommand{\seclabel}[1]{\label{sec:#1}}


%% file: abstract.tex


\begin{abstract}

Longitudinal passive sensing studies for health and behavior outcomes often have missing and incomplete data. Handling missing data effectively is thus a critical data processing and modeling step. Our formative interviews with researchers working in longitudinal health and behavior passive sensing revealed a recurring theme: most researchers consider imputation a low-priority step in their analysis and inference pipeline, opting to use simple and off-the-shelf imputation strategies without comprehensively evaluating its impact on study outcomes. Through this paper, we call attention to the importance of imputation. Using publicly available passive sensing datasets for depression, we show that prioritizing imputation can significantly impact the study outcomes -- with our proposed imputation strategies resulting in up to 31\% improvement in AUROC to predict depression over the original imputation strategy. We conclude by discussing the challenges and opportunities with effective imputation in longitudinal sensing studies.
\end{abstract}

%% file: headings.tex


\ifsubmit
    \relax
\else
    \par\noindent \textcolor{red}{\textbf{DRAFT}: \today\ -- \currenttime}
    \pagestyle{fancy}
    \lhead{DRAFT in preparation}
    \rhead{version: \today\ -- \currenttime}
    \chead{}
    \lfoot{}
    \rfoot{}
\fi

%% file: body.tex


\section{Introduction} 
\seclabel{introduction}

The ubiquitous presence of smartphones and wearables has enabled researchers to collect rich and high-resolution data on individuals' daily activities, interactions, and physiological responses over extended periods~\cite{cornet2018systematic}. This widespread adoption has profoundly influenced the landscape of mobile health research, leading to a surge in longitudinal studies passively examining participants' behavior.
These studies capture multiple dimensions of a person's life -- mental health~\cite{xu2022globem, wang2014studentlife, wang2016crosscheck,mohs2000longitudinal}, academic performance~\cite{huang2015academic,marsh1997causal, das2020leveraging}, workplaces~\cite{mirjafari2019differentiating, dasswain2019multisensor, DasSwain2019FitRoutine}, medical symptoms tracking~\cite{reyna2020early,gagnon2022woods}, sleep wellness~\cite{zhang2021relationship,mork2012sleep}, and physical activity~\cite{kunzler2019exploring}. 
Data from longitudinal studies can reveal patterns, fluctuations, and trends that might not be apparent in cross-sectional studies, where data is collected at a single point in time. Despite the potential of passive sensing studies in monitoring longitudinal trajectories, ensuring data completeness remains a persistent challenge ~\cite{laird1988missing,bell2014practical}.


Data missingness in longitudinal studies can occur due to several factors, including participant's behaviors leading to low compliance and  technical problems with the sensing apps/devices~\cite{vhaduri2017design, moller2013investigating}. Many longitudinal studies incentivize participants to be compliant through monetary rewards~\cite{campbell2023patient} or altruistic motives~\cite{kramer:step-goals}. While incentives can attract participants, they do not guarantee full compliance~\cite{campbell2023patient}. Participants may become disengaged over time or fail to adhere to data collection protocols, leading to gaps in the collected data~\cite{young2006attrition}. For instance, participants may forget to charge their smartphones or wearables regularly, resulting in periods of data loss when devices power off due to low battery levels. Similarly, participants may inadvertently switch off sensors or disable data collection features, leading to gaps in the recorded data.
Moreover, the appeal of incentives may attract participants who are less motivated to contribute accurate or reliable data, further exacerbating data completeness issues~\cite{huang2015insufficient}. Even when participants adhere to data collection protocols, technical errors or bugs in data collection processes can compromise data completeness. For example, software bugs or compatibility issues may cause data loss or corruption during transmission or storage. Furthermore, these studies are often conducted in the wild, which entails many unforeseen circumstances that can add to data missingness.

Missing data in longitudinal studies, whether due to technical challenges or participant-related factors, could present an incomplete view of behavioral patterns and researchers may fail to capture the longitudinal shifts or developmental trajectories. Thus, compromising the reliability and accuracy of the scientific insights researchers may seek to derive from the data~\cite{newman2014missing, barnett2018beyond}. 
For instance, consider a study tracking physical activity patterns over a span of several months and suggesting interventions to promote a healthy lifestyle. Periodically missing physical activity data, however, could lead to an underestimation of activity levels, potentially impacting conclusions about the effectiveness of interventions or lifestyle changes.
The presence of missing data may also introduce bias into the study results, as the observed sample may not be representative of a person's behavior. Thus, missing data, if not handled properly, can not only impact behavioral data analysis but can also be potentially detrimental to participants if these analyses are used for outcome predictions and interventions.
Researchers are increasingly using sophisticated deep learning and machine learning methods to model and predict various health outcomes with passive sensing data. These approaches, particularly the ones with neural networks (e.g., LSTMs, FNNs, etc.) require complete data and necessitate researchers to handle missingness before feeding the data into such models. 
To reduce data missingness, researchers often employ protocols to monitor data collection constantly and check with participants through emails and phone calls if they see missing data from the participant~\cite{christensen2015effect, triplet2017mail}. Despite this manual and labor-intensive process, data missingness is unavoidable in longitudinal research~\cite{campbell2023patient}, with recent studies reporting missing data rates from 20\% of over 50\% for certain features and data types~\cite{nepal2024capturing, xu2022globem, xu2023globem, rashid2020predicting}. 

Addressing data missingness is a complex process that involves making decisions about whether to drop missing data, use imputation techniques, or apply a combination of both approaches. Researchers’ decisions can directly affect the data they use for subsequent analyses and model building. Despite the importance of this process, it appears to be often overlooked in passive sensing research. This observation stems from the fact that prior works in passive sensing studies often tend to focus on novel outcomes or novel machine learning and modeling methods, with relatively limited discussion on handling missing data or imputation strategies. As a first step, we wanted to understand whether data missingness remains \textit{understudied} or simply \textit{underreported}. Thus, to gain a deeper insight into researchers' decision-making processes, practices, and challenges with handling data missingness, we conducted formative semi-structured interviews with thirteen researchers with experience in working with longitudinal passive sensing health data. 

The researchers we interviewed revealed that they often considered addressing data missingness and imputation as a lower priority than other aspects of the study, such as model or algorithm development and subsequent post-study analyses. When imputing data, researchers tend to use simple or off-the-self strategies without comprehensively evaluating the validity of the imputed data or its impact on the study outcomes. Through this paper, we draw attention to the critical role of selecting appropriate and robust imputation strategies in health and behavioral longitudinal studies. To this end, we present a case study using the GLOBEM platform and its associated publicly available datasets for depression sensing and detection~\cite{xu2022globem, xu2023globem}. Our evaluations reveal that different imputation strategies can significantly influence study outcomes without any modifications to the data processing and modeling pipeline. Additionally, we present a novel Autoencoder-based imputation strategy, highlighting that investing time and effort in the development of imputation strategies can lead to higher predictive performance of models.\hey{check this line} We showed that our Autoencoder-based imputation strategy could achieve \textit{up to a 31\% improvement in AUROC} for within-person future depression prediction compared to the imputation approach in the original GLOBEM pipeline~\cite{xu2023globem}.

\medskip
\noindent
In this paper, we make three important \textbf{contributions}: 
\begin{itemize}
    \item  We provide insights on the current decision-making process and challenges of researchers when addressing data missingness in longitudinal health and behavior sensing data.

    \item We present a comparison of commonly used simple and off-the-shelf imputation strategies and demonstrate importance for investing in the development of imputation strategies using GLOBEM datasets. 
    
    \item We discuss challenges with data missingness and the potential opportunities for effective imputation strategies. 
\end{itemize}

Our code for data missingness analysis and imputation strategies are publicly available at (GitHub link)\footnote{anonymized for submission} for other researchers to explore and implement imputation strategies on their datasets. We emphasize that our intention is not to undermine the importance of other aspects of the data pipeline but to underscore the role of effective imputation and to encourage further discussion and attention to it in future research.

\section{Related work}
\label{sec:related-work}

Longitudinal studies provide a comprehensive understanding of the underlying behavioral, affective, and mental processes by continuously tracking changes in behavior over an extended period. Researchers in the past have conducted longitudinal studies to track a wide spectrum of human behaviors~\cite{xu2022globem, wang2014studentlife, marsh1997causal, zhang2021relationship, nepal2024capturing, wang2018sensing, grunerbl2012towards, das2022semantic}. However, conducting long-running longitudinal studies is a demanding task requiring significant time, resources, and funding, with ensuring data completeness being a big challenge.

\subsection{Missing Data in Longitudinal Studies}

Missing data in longitudinal studies has been attributed to low participant compliance, attrition, and technical issues. Participants may drop out of the study over time for various reasons, such as relocation, loss of interest, or inability to continue participation~\cite{little1995modeling, wolke2009selective, philipson2008comparative}. Studies based in schools and universities have reflected attrition due to students moving schools, being absent, or refusing consent~\cite{bonell2019effects, mcdonald2017implications}. The demographics of participants can play a crucial role in attrition rates. Young et al.~\cite{young2006attrition} showed that younger individuals have a much higher tendency to drop out of studies as compared to middle-aged or older individuals. Past works have tried simulating attrition to account for it in designing longitudinal studies~\cite{kristman2005methods,ibrahim2009missing}. Researchers have also outlined guidelines to reduce attrition rates~\cite{leeuw2005dropout, fumagalli2013experiments}.
Even in studies with low attrition rates, data missingness still remains a persistent problem as participants may choose not to respond to surveys or adhere to data collection protocols. Longitudinal studies collecting information on sensitive topics have shown higher rates of non-response or missing data due to participant discomfort or reluctance to disclose~\cite{loxton2019longitudinal,griesler2008adolescents}. Passive sensing longitudinal studies rely on mobile phones, wearables, and sensors to collect data, which leads to missing data stemming from technical and connectivity issues with devices~\cite{xu2022globem, wang2014studentlife, chow2017using, bahr2022missing}.

Typically, to reduce data missingness researchers  continuously monitor data collection streams and reach out to participants with missing data. Triplet et al.~\cite{triplet2017mail} and Christensen et al.~\cite{christensen2015effect} sent reminders to participants who did not complete self-reported surveys within 12 hours. Sano et al.~\cite{sano2018identifying} prompted participants to revise their answers if missing or inaccurate values were detected. Despite these preventive and labor-intensive mechanisms in place, data missingness is often inevitable in longitudinal studies. In fact, in a longitudinal study on opioid use, Campbell et al.~\cite{campbell2023patient}, reported, on average, 70\% survey completion rates even after constantly being in touch with patients. There are multiple ways researchers have handled missing data in practice , especially while dealing with smartphone and wearable sensor data - viz. (a) dropping features or days where a threshold of feature availability is not met, (b) completely removing data for participants if the missing rate is above a certain threshold (c) employing data imputation strategies by making estimations for missing values using observed data in the study.

\subsection{Dropping Incomplete Data}



Researchers often drop features, days, or participants' entire data to ensure data quality. For the StudentLife study, Wang et al.~\cite{wang2014studentlife} removed a student's data if their phone data was missing due to power issues or the phone being left at the dorm. 
Similarly, Obuchi et al.~\cite{obuchi2020predicting} removed participants who had less than 18 hours per day and less than 14 days of data during the term. Likewise, Wang et al.~\cite{wang2018sensing} removed participants with less than 19 hrs of data on a day and less than 7 days of usable data. They mentioned that only 159 out of the 646 participants satisfied their data inclusion criteria and were included in the analysis. In a multimodal smartphone sensing study~\cite{assi2023complex} on complex daily activity recognition, the authors dropped features from sensors that were missing more than 70\% of times. A study on drinking behavior sensing ~\cite{santani2018drinksense} used three criteria to select user data - the participant had responded to either the drink or the forgotten drink survey at least once during a given night; b) the participant had at least one data sample for any sensor data type during the night; and c) the participant had at least one stay-point based on location sensor logs.  Huckins et al.~\cite{huckins2019fusing} chose a threshold of at least 20 days of available data to keep participants in a study for depression assessment using smart phone data. Similarly, In a study by Grunerbl et al.,~\cite{grunerbl2012towards} on passive monitoring of bipolar disorder, data from only four out of ten patients was considered appropriate for analysis due to attrition and data inconsistencies. Some studies in the past had to remove data due to health conditions~\cite{salthouse2014selectivity}, privacy~\cite{saha2019imputing}, and even the death of the participants~\cite{teodorczuk2007white}. Even after excluding participants' data with high levels of missing information, data missingness still persists in the datasets for participants' data below the data removal thresholds. To address this, researchers use \textit{data imputation strategies} to estimate missing data using available data of the participants.

\subsection{Imputing Incomplete Data}
Collecting passive sensing data on health and behavior is a task that requires significant effort and time \cite{hernandez2017data}. Discarding all the data records with missing values often leads to datasets that are too small to be useful, making this approach an impractical solution. Moreover, most of these studies build Machine Learning and Deep Learning models that often require complete input data feature vectors for training and inference, thus necessitating the imputation of missing data.
For mobile and wearable sensing data, researchers have used simple statistical imputation strategies (like mean, median, mode, and LOCF) and statistical interpolation methods. Xu et al. \cite{xu2023globem} used mean and median imputation strategies for filling up missing data in the GLOBEM Platform. Similarly, Chow et al. \cite{chow2017using} used LOCF to estimate missing values in global positioning system (GPS) data, whereas Burns et al.~\cite{burns2011harnessing} used mixed models to get a best-fit line for ``estimated GPS'' coordinates when the phone cannot obtain GPS readings due to connectivity. In cStress~\cite{hovsepian2015cstress} study, authors used cubic Hermite splines to interpolate high-frequency Electrocardiogram and respiration data. Likewise, DaSilva et al.~\cite{dasilva2019correlates} imputed the missing MPSM (Mobile Photographic Stress Meter) scores using Kalman smoothing~\cite{aravkin2017generalized}. Researchers have also used kNN-based strategies for sensor data ~\cite{assi2023complex, sarker2016finding} and have also built machine-learning models to impute missing passively sensed data . Saha et al. ~\cite{saha2019imputing} imputed social media features using sensor data from phones and wearables by building Linear Regression, Gradient Boosted Regression \& Multilayer Perceptron models. Sano et al. ~\cite{sano2018identifying} built automated classifiers to detect noisy and incomplete data and asked participants to revise them, whereas Choube et al. ~\cite{choube2024sesame} suggested using Large Language Models to simulate incomplete survey responses.

Previous studies have employed multiple imputation strategies; however, they did not assess or discuss the consequences of their chosen imputation strategies except the work by Rashid et al.~\cite{rashid2020predicting}. The authors collected a five-week passive sensing dataset and built ML models to predict social anxiety. They compared model performances when LOCF, kNN, Multiple Imputations by Chained Equations (MICE) \cite{van2011mice}, and Matrix Completion \cite{mazumder2010spectral} imputation were used. On average, LOCF, kNN, MICE, and Matrix Completion resulted in the reduction of RMSE by 11\%, 13\%, 17\%, and 20\%, respectively, when compared to the non-imputed dataset. Rashid et al. presented a comparison of strategies but did not discuss the decision process behind choosing these strategies and the associated challenges with imputation. In a workshop paper, Bhattacharya et al. ~\cite{bhattacharya2024imputation} compared multiple imputation strategies for predicting depression and tracing within-person transitions in depression. 




In this paper, we take a more human-centered approach, starting with understanding the researcher's current practices, perspectives, and challenges involved with handling data missingness and choosing an imputation strategy through semi-structured interviews. We then present a case study with six datasets, demonstrating that investing time in imputation strategy development and selecting the right imputation strategy can play a pivotal role in affecting study outcomes.

\section{Formative Study: Handling Data Missingness in Practice}
\label{sec:formative-study}

We conducted qualitative interviews with researchers to gain insights into their decision-making process and challenges they encounter while handling missing data and selecting an imputation strategy in their studies.

\subsection{Study Method}

We chose a convenience sampling strategy and reached out to 13 researchers (faculty members (3), postdoctoral scholars (2), research scientists (2), and doctoral students (6)) experienced in working with longitudinal passive sensing data ($\mu = 6.23$ years, $\sigma = 3.72$ years) for health and behavior outcomes. Eleven researchers expressed that they are professional in dealing with longitudinal health and behavioral datasets, while two researchers said they are somewhat professional. Some of these researchers are also authors of popular and well-received datasets and publications in this field.
We conducted 30-minute interview sessions with participants over Zoom. After receiving their verbal consent, we recorded the meeting. We used transcripts from the meeting to generate our insights. Participants also completed a short survey on demographic information at the end of the interview session. To thank participants for their time, we compensated them with a $\$15$ Amazon gift card. Our interview protocol was approved by our university's Institutional Review Board (IRB).

We used MS Word's automatic transcribing tool to transcribe our interviews. Two members of our research team independently coded four interviews to develop the initial codes. Both the coders carefully reviewed the transcripts and employed the Open Coding approach \cite{corbin1990grounded}. After independently coding four interviews, the coders met to discuss their codes -- merging codes and resolving conflicts through consensus. After this, the first author coded the rest of the interviews. 

\subsection{Interview Insights}
Our interviews with researchers revealed interesting insights into potential causes of missingness in studies, their decision-making process in handling missingness, and the associated challenges.


The researchers confirmed that data missingness is a highly prevalent issue, especially in free-living studies. As P11 said, \textit{``I have come to accept that it's part of life and most of the pipelines that we have developed take this into account and are built on the premise that there will be missing data"}.
Researchers attributed technical issues with passive sensing devices and data collection applications (or servers) as a prominent cause. P3 mentioned, \textit{`` A reason (for missingness) is network issues or the mobile app that you have developed has a bug (like) memory out of bounds or app is going crazy.’’} 
Researchers also mentioned participants' low compliance as a major reason, highlighting that participants may switch off their phones or stop wearing smartwatches due to privacy concerns, personal preferences, or discomfort. 
P4 noted that \textit{``A lot of times users may turn off their location for privacy reasons or maybe for battery saver reasons.''} Some researchers studying specialized populations, schizophrenia patients, opioid recovery patients, and older adults mentioned that low motivation or technical skills in specialized populations might lead to lower compliance. P5 explained,  \textit{``One reason that is not highlighted as often is that sometimes people don't even know how to operate phones, like people who are socially less functioning or on severe spectrum. It's hard to educate them and maintain the quality of incoming data.''}
Researchers also explained that understanding the reasons for missing data might be challenging due to a lack of context about participants' lives, not having research coordinators available to regularly reach out to participants for clarification, and/or not having collected the datasets themselves. P6 mentioned that \textit{`` Missingness is a large issue, and I think we spend a lot of time trying to understand why missingness occurs because often it's not quite clear to us, especially because a lot of times if the focus is on the machine learning end, we might not have been a part of the larger clinical study that would collect the data.''}

To deal with data missingness, the first decision researchers must make is determining which portion of the data to drop and which portion to impute. Data missingness percentage plays an important role in this decision. When the percentage of missing data is low, researchers often prefer to drop the records with missing data, as the remaining dataset would still be large enough to yield meaningful conclusions. P1 noted, \textit{“If the missingness isn’t too much, I’ll drop the data and move on. But if more than 25\% is missing, I’ll try to impute it.”} Researchers also discard parts of participants' data that do not meet specific thresholds. These thresholds vary among researchers -- some treat them as hyperparameters to experiment with, while others select values they deem appropriate. P2 remarked, \textit{``Sometimes I have a somewhat random threshold, like, if more than half of data is missing, then it doesn’t make sense for us to impute \ldots\ It is a hyperparameter that can be tuned, but I have never experimented with it.''}
Researchers also expressed that they are more comfortable in discarding data that is easier to collect. P9 noted, \textit{``For datasets that get continuously generated, it's okay to not invest time in imputation, and you can drop data without much information loss, but in some cases removing rows can lead to information loss, for example, experience sampling data, the imputation would make sense.''} Additionally, researchers highlighted that phone-sensing data is generally more difficult to collect. They also felt more at ease in dropping data from in-the-lab studies compared to in-the-wild studies. 

Some researchers expressed skepticism about adding synthetic data to the collected ground truth data. P7 explained, \textit{``I am gently cautious of simulating or imputing data. Not that I'm morally opposed to it, but I'm like a lot of things need to be validated before I do it\ldots I think I'm actually more on the side of if you know that data is missing from just a portion of it, I would just not use that portion.''} Other researchers were more comfortable imputing input sensor data but were hesitant to do so for target outcome data, particularly self-reported data.  For instance, P6 mentioned, \textit{“I have a lot of discomfort around filling data for a target outcome sometimes especially self-reported data”}. Some researchers expressed concern when imputing highly sensitive data streams like Electrocardiogram data.
 

The next decision researchers have to make is to decide which imputation strategies to employ. Most researchers mentioned using simple statistical strategies like mean, median, and forward-fill. Some researchers mentioned using off-the-shelf strategies like Multiple Imputation by Chained Equations (MICE) and k-Nearest Neighbour (kNN) imputation. The nature of data played a role in choosing the algorithm and P1 stated \textit{`` It depends on the modality of the sensing stream, if they're not very granular activities, I would forward fill them, if data is continuous measurements, I would use mean or median.''}. Often, limited bandwidth, deadlines, and prioritizing other parts of the study led to researchers using simple strategies. P2 explained, \textit{``If the deadline is pushing and we have limited bandwidth, we pick mean or median, that is a most straightforward way, and then go with it.''}. Similarly, P6 remarked \textit{``I spend more time thinking about it (imputation) in general and less time like executing well in any individual study}''. Some researchers preferred using well-established methods as they felt it would be easier to convince paper reviewers.   

Some researchers thought of choosing an imputation strategy akin to choosing \textit{hyperparameter} or a \textit{hit and trial} process. They expressed that although they try some simple and off-the-shelf strategies, they do not comprehensively evaluate the strategies before choosing the strategy for their datasets. P6 mentioned, \textit{"At least in my lab, there was no set rule (choosing imputation). It was just ad hoc. Like whatever works \textit{hit and trial}. We just use that.''}. Similarly, P9 explained, ``As a process we generally do (compare strategies) but it's not being that systematic \ldots\ I have tried simple experiments like deletion vs imputing with mean or median.''


While most researchers recognized data imputation as an important process in the data pipeline, acknowledging its direct impact on the data, they agreed that historically they have given less attention to this step in their work. For instance, when asked to rate the significance of data imputation in longitudinal studies, \textit{P2 responded, ``I think it's gonna be 7 or 8. It's very critical \ldots\ but in terms of my actual practice, maybe I just do it at four or five.'}'
We asked researchers about their perceptions and experiences on whether the choice of imputation strategy could impact their study outcomes. Some researchers stated that they doubt there will be a significant influence; some felt it was crucial, while others were uncertain. For example, P1 said,   ``\textit{I haven't seen much improvement (on study outcomes). The main reason for that is because the data we collect in the wild, it's already very noisy, and I feel like imputing based on that does not really create a signal that is actually useful.''} while P12 remarked, \textit{"I have observed that (significant influence), but at the same time, there's no universal pattern."}

On probing further about how they decided on handling missingness or imputation strategies, most of them mentioned referring to previous similar papers in their field. Some researchers, however, noted that these papers often lack detailed discussions on imputation. As P4 explained, ``I think I definitely reference the previous paper in my field \ldots\ Sometimes they don't report. Actually, a lot of times.'' Some researchers followed the suggestions provided by their collaborators, while others relied on their intuition. Many expressed that having a more comprehensive guideline for handling missing data would be beneficial. P2 shared, \textit{``I make decisions based on my intuition and pre-experience, but I wish we had a central guideline.''}



Our findings on researchers' decision-making processes and the challenges they face when addressing data missingness reinforced our assumption that the imputation step often receives lower priority and effort compared to other aspects of the study. Researchers' current practices often lacked a systematic approach to choosing imputation strategies and were more ad-hoc and hit-and-trial based. Our results indicate that handling missing data and selecting imputation strategies is not simply underreported but also understudied. Additionally, some researchers mentioned not observing different imputation strategies to have a significant impact on the study outcomes. We hypothesize that their observation could be because they rely on overly simplistic measures for imputation, like mean, median, and forward fill, without systematically comparing these simple strategies with other more complex imputation strategies.  In the next section, through a case study using six publicly available GLOBEM datasets, we illustrate that evaluating and selecting the appropriate imputation strategies can significantly improve study outcomes. 

\section{GLOBEM Datasets and Platform}
\label{sec:the-globem-datasets-and-platform}

As part of GLOBEM, Xu~et~al.~\cite{xu2022globem} have made their four datasets publicly available. They collected these four datasets (INS-W\_1, INS-W\_2, INS-W\_3, and INS-W\_4) across four years, from 2018 to 2021, at a university. In addition to these datasets, Xu~et~al.\ also compared their approaches with two additional datasets (INS-D\_1 and INS-D\_2), collected at a second university from 2018 to 2019~\cite{xu2023globem}, which they graciously shared with us for our work. For the purposes of this paper, we refer to these six datasets collectively as GLOBEM datasets.
Each dataset included data from the Spring quarter (approximately 10 weeks) for each participant to control for seasonal effects.
These datasets include a broad spectrum of features that the authors collected continuously in the background, including call, sleep, location, steps, phone screen activity, and Bluetooth. Participants also completed two comprehensive surveys on health behaviors, social well-being, emotional states, and mental health during the start and end of the study. In addition to pre- and post-study surveys, participants answered Ecological Momentary Assessment (EMA) surveys every week (PHQ-4, PSS-4, PANAS) measuring depression and stress. In the GLOBEM datasets, each day is split into four epochs (time intervals of 6 hours each) -- morning, afternoon, evening, and night. The feature data collected over a particular epoch is aggregated as the cumulative data of the entire epoch. Thus, GLOBEM datasets contain features aggregated over the entire day and four epochs of the day. For additional details on the different datasets, please refer to the original GLOBEM papers~\cite{xu2022globem, xu2023globem}. 

\begin{table}[h]
\small
\caption{Average data availability (\%) of feature data for the Reorder and Chikersal feature sets for GLOBEM datasets}
\Description{Average data availability (\%) of feature data for the Reorder and Chikersal algorithms for GLOBEM datasets. We see that there is high missingness in Reorder feature set compared to Chikersal feature set}
\begin{tabular}{|l|llllll|}
\hline
             & \multicolumn{6}{c|}{\textbf{Datasets}}                                                                                                                 \\ \hline
\textbf{Algorithms}   & \multicolumn{1}{l|}{W\_1}  & \multicolumn{1}{l|}{W\_2}  & \multicolumn{1}{l|}{W\_3}  & \multicolumn{1}{l|}{W\_4}  & \multicolumn{1}{l|}{D\_1}  & D\_2  \\ \hline
Reorder   & \multicolumn{1}{l|}{33.71} & \multicolumn{1}{l|}{35.67} & \multicolumn{1}{l|}{31.07} & \multicolumn{1}{l|}{35.43} & \multicolumn{1}{l|}{38.1}  & 40.24 \\ \hline
Chikersal & \multicolumn{1}{l|}{60.1}  & \multicolumn{1}{l|}{61.7}  & \multicolumn{1}{l|}{59.7}  & \multicolumn{1}{l|}{65.5}  & \multicolumn{1}{l|}{71.76} & 75.32 \\ \hline
\end{tabular}
\smallskip

\label{tab:data-missingness}
\end{table}


The authors also provided an open-source benchmarking platform, ``GLOBEM Platform''~\cite{xu2023globem} with 19 depression detection algorithms and multiple evaluation tasks. The GLOBEM platform supports the development and evaluation of different longitudinal behavior modeling methods. For our analyses and evaluations, we chose best-performing machine learning algorithm \textbf{``Chikersal''}~\cite{chikersal2021detecting} and deep learning-based algorithm \textbf{``Reorder''}~\cite{xu2023globem} available in GLOBEM Platform.

\subsection{Data Availability in GLOBEM}
\label{subsec:data-availability-across-datasets}

The GLOBEM datasets are a tremendous resource for researchers to advance the field of behavioral modeling and depression prediction with passive sensing data. These datasets, however, suffer from considerably high data missingness. We report the overall data availability metrics for features used by Reorder and Chikersal algorithms in Table~\ref{tab:data-missingness}. We calculate data availability as the percentage of non-empty cells in the dataset. 
The feature sets for Chikersal and Reorder Algorithms are quite different, with just eight features in common. The Chikersal feature set has considerably less missingness than the Reorder feature set. 
The open-source code of the GLOBEM Platform and the data missingness in the datasets make them ideal datasets to elucidate (1) the importance of data imputation as a preprocessing step and (2) the challenges associated with imputation, particularly with varying levels of data missingness for the Chikersal and Reorder algorithms.


As a first step, we evaluated if the data is Missing Completely at Random (MCAR), using the Little's MCAR Test~\cite{little1988test} provided by the R package \texttt{missMDA}. Some prior works on longitudinal passive sensing data conducted the MCAR test and reported $p > 0.05$ thus concluding that the data was likely MCAR~\cite{nepal2024capturing, rashid2020predicting}. However, considering they only reported \textit{one} $p$-value it appears that prior works may have conducted one MCAR test for the entire dataset. Given the longitudinal nature of the data and potential within-participant variability, we argue that it is more appropriate to conduct individual MCAR tests for each participant. Thus, we conducted Little's MCAR tests for each participants' data with Benjamini–Hochberg correction (to account for multiple testing), and found $p<0.05$ for over 70\% of the participants, thus suggesting that the data might not be MCAR.

\begin{figure}[h]
  \centering
  \includegraphics[width=0.8\linewidth]{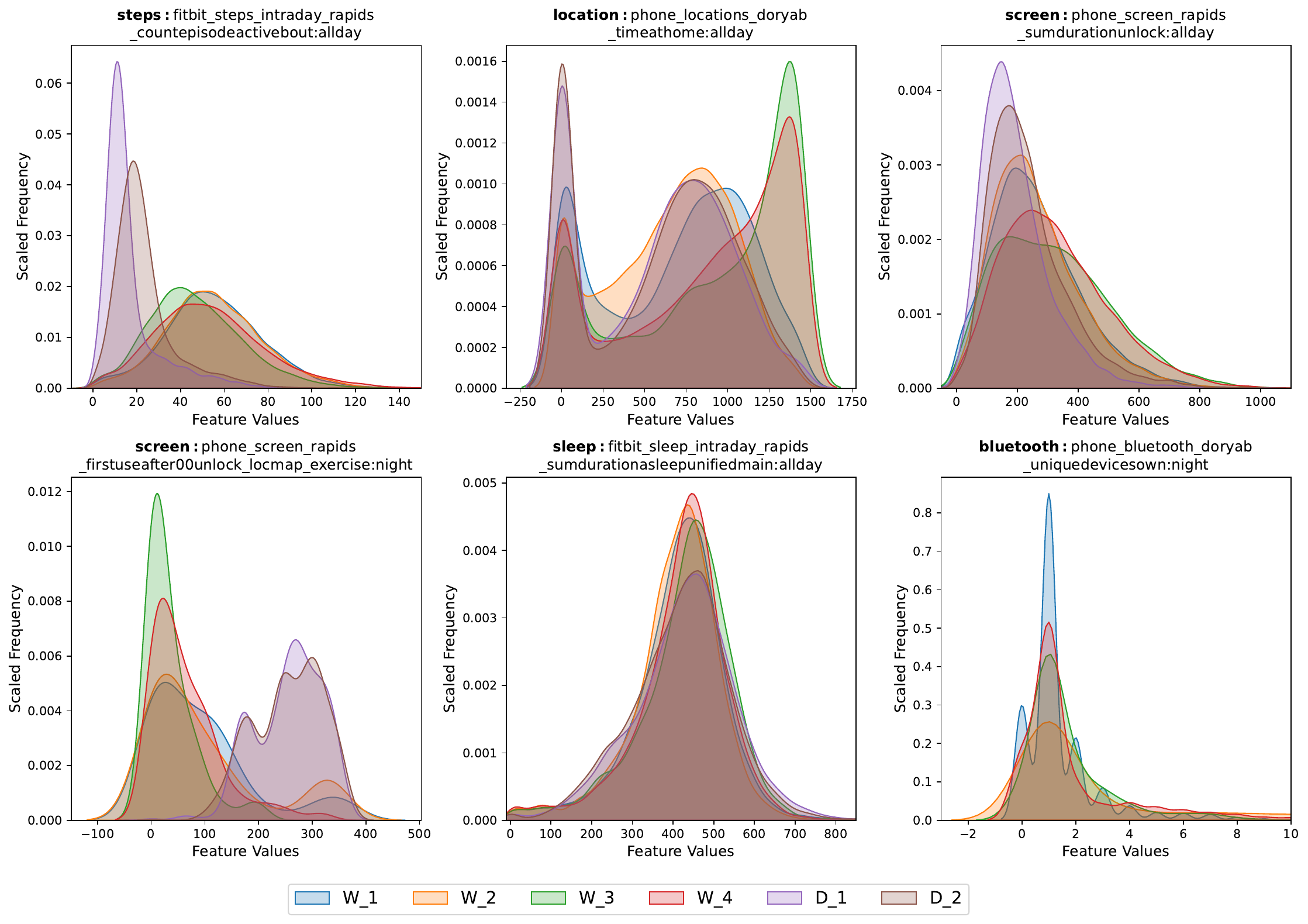}
  \Description{Density Distribution of 6 features across six datasets. Sleep features are similarly distributed in most datasets, while Step features are different across Datasets, most noticeably in INS-W versus INS-D.}
  \caption{Density distribution of some features from different feature categories (steps, location, screen, sleep, and bluetooth for six GLOBEM datasets. The density distributions of most features tend to vary across different datasets. The features from sleep category are more similar as expected.}
  \label{fig:density_dist}
\end{figure}

\subsection{Variability in GLOBEM Datasets}
\label{subsec:feature-data-dispersion}



We examined the distributions of the key features across the six GLOBEM datasets. We visualized the distribution of feature values using kernel density plots to identify differences in the shape, skewness, and tail behavior of feature distributions. We present density plots of some of the representative features from different categories in Figure \ref{fig:density_dist}. The density plots show significant variations in feature distributions across datasets. For instance, location features like the number of locations visited per day, as well as time spent in locations, i.e., work, home, etc., were significantly different in W\_3 and W\_4 as compared to data collected in W\_1 and W\_2, possibly due to enforced restrictions on outdoor movement during COVID-19 years. 


The observed differences in feature data dispersion and distribution suggest that datasets are diverse, and so are individual participants. Thus, although sensor data features collected as part of the GLOBEM framework are consistent across all six datasets, the diversity of datasets adds to the generalizability as well as the robustness of the analyses and results in subsequent sections.

\section{Methodology}

In our formative interviews, researchers mentioned focusing on different study outcomes, i.e, researchers conduct longitudinal passive sensing studies with broadly two objectives: first, to observe patterns, describing associations, and clarifying/discovering phenomena in longitudinal data using statistical methods without
involving model construction or outcome prediction (\textbf{Reconstruction Task}). Second, to track and predict certain outcomes related to human activity, behavior, or health.
In these studies, researchers collect longitudinal data and then build machine learning, deep learning, and statistical
models to predict outcomes (\textbf{Prediction Task}). Hence, in this work, we chose to analyze imputation strategies for both study outcomes.

\subsection{Reconstruction Task}

If the goal of a longitudinal study is to assess correlations between various behavioral features, then imputation should be able to fill in the values of a feature close to its original distribution. To compare different imputation strategies, we can remove some portion of available data and then use the strategies to impute the removed data. We can then assess the closeness of imputed values with actual values using metrics like root mean squared error (RMSE) or mean absolute error (MAE). 

Consider, we remove $n$ values from the dataset $D$ using some amputation strategy ($A$). Let the removed indices be $\{(i_m,j_m ),\text{ }m =  1 \cdots n\}$. Then, we use the imputation strategy '$I$' to impute and get the imputed dataset $\hat{D}$. we can compute reconstruction RMSE (r-RMSE).

$$\text{\textit{r}-}RMSE_I = \sqrt{\frac{1}{n} \sum_{m=1}^{n}(D_{(i_m,j_m)} - \hat{D}_{(i_m,j_m)})^2}$$


We explore following amputation Strategies ($A$):

\begin{itemize}
    \item MCAR: This strategy amputes data completely at random. To simulate empirical imputation error for $r\%$ missingness, we iterate through each participant's data and remove $r\%$ of feature data per participant per feature \textit{randomly} 

    \item MNAR: This strategy removes data assuming the data is missing not at random. For MNAR amputation, we try two ways of data removal: 
    \begin{itemize}
    \item MNAR (i): Higher than $(50 -  k_{i}/2)^{th}$ percentile and lower than $(50 +  k_{i}/2)^{th}$ percentile of the target feature variable. 
    \item MNAR (ii): Lower than $(k_{ii}/2)^{th}$ percentile and higher than $[100-(k_{ii}/2)]^{th}$ percentile of the target feature variable~\cite{kazijevs2023deep}.
    \end{itemize}
    For a fair comparison with the MCAR method, we chose $k_{i}$ and $k_{ii}$ respectively such that it accounts for an overall $\sim r\%$ missing rate.
\end{itemize}

We chose $r = 10$ \footnote{ We ran experiments with $r \in [10, 20, 50]$ and found that r-RMSE kept decreasing with lower amputation percentage, but trends were similar. Hence, we only report $r = 10$.} for our analyses. The lower the r-RMSE, the better the imputation strategy for the reconstruction of feature values.  We used Reorder and Chikersal feature sets for our reconstruction loss comparisons.

\subsection{Prediction Task}

If the goal is to build systems and models to predict health and behavioral outcomes, then imputation should help ensure high predictive performance of the outcomes. Researchers often measure predictive performance with metrics like F1 score, balanced accuracy, and area under the receiver operating characteristic curve (AUC).
Therefore, to compare the success of different imputation strategies, we need to compare the relative boost in predictive performance by these strategies. In this study, for prediction task, we focus on the task of predicting depression labels for future weeks of participants using sensor and survey data from their past weeks. This task was first defined in the original paper \cite{xu2022globem} and is available as the ``within-user'' task in the GLOBEM platform. We used two prediction model algorithms from the original paper -- Reorder and Chikersal. The prediction models were trained on data from 80\% of the weeks and then used to predict depression labels for the remaining 20\% of weeks. We use the same evaluation metrics (balanced AUROC and balanced accuracy) as used by the original study.

Additionally, as longitudinal studies can run for extended periods spanning months or even years, researchers might want to analyze
data and build models during multiple checkpoints in the study. Thus, we wanted to explore data imputation trends and
performances of imputation strategies in real-time settings. We simulated a real-time inductive imputation scenario
where we start predicting depression labels starting at the end of week 3 until the end of week 10.  We use $n-1$ weeks data to impute and then predict depression labels for the $n^{th}$ week.

\subsection {Imputation Strategies}

For the tasks mentioned above, we evaluate various commonly employed and off-the-shelf imputation strategies such as Median, Multiple Imputations by Chained Equations (MICE)~\cite{van2011mice}, k-Nearest Neighbour (kNN)~\cite{troyanskaya2001missing}, and Matrix Completion~\cite{mazumder2010spectral}. We used the \texttt{sklearn} library's  \texttt{IterativeImputer} and  \texttt{KNNImputer} for MICE and KNN strategies, respectively. For Matrix Completion we used \texttt{fancyimpute} library's \texttt{SoftImpute} function for our experiments. We also developed two \textit{novel} algorithms:

 \subsubsection{Bounded kNN}: This variant of kNN allows a choice of a lower bound ($l$) on a number of neighbors to be present for imputation to happen. If there are less than $l$ neighbors present, we do not impute the missing value. We also constrain existing feature values within the $5^{th}$ and $95^{th}$ percentiles of their respective distributions while calculating the distance between neighbors to ensure that outlier values do not impact our neighbor selection. For our experiments, we set the lower bound ($l$) as $2$ and the upper bound ($u$) as $6$ based on empirical evaluations. Throughout the rest of the paper, we refer to this variant of kNN as bounded kNN, while the standard off-the-shelf kNN is called Simple kNN.

\subsubsection{Autoencoder based Algorithm:}
Autoencoders are a type of artificial neural network that learns to encode input data into a lower-dimensional representation and then decode it back to its original form. It consists of an encoder network that compresses the input data into a latent space representation and a decoder network that reconstructs the original input from this representation.

Mathematically,
$$f_{\omega,b}(x) \approx x$$

Where $x$ is the input, $\omega$ and $b$ are parameters of the model, and $f$ is the learned function. Using Autoencoders as data imputers involves the following steps:
\begin{itemize}
    \item Let $D_M$ be a dataset containing missing values. We use a simple initial imputation strategy to fill in the missing values in $D_M$. This step is essential as an autoencoder neural network requires input data to be complete, and Median, Mean, and kNN strategies can be used as initial imputation strategies.
    
    $$D_C = InitialImputation(D_M)$$
    where $D_C$ is the completed dataset and $InitialImputation$ is the initial imputation strategy used.
    
    \item Train the autoencoder using the completed dataset ($D_C$) where loss is only calculated for data points that were originally present in the dataset before initial imputation.
    
    \item After the autoencoder is trained, we pass the completed dataset $D_C$ and get output as $\tilde{D}_C$. Note that $\tilde{D}_C$ will also be a complete dataset.
    
    \item In $D_M$, fill the missing values using values present in $\tilde{D}_C$ at respective indices. 
\end{itemize}

We implemented the Autoencoder imputer using \texttt{PyTorch}. We used one hidden layer with a dimension of 20.\footnote{We also experimented with additional hidden layers and custom loss functions, but they performed at par with the current approach; hence, for brevity, we only discuss a single autoencoder implementation.} With the autoencoders, we experimented with Median (Autoencoder-Median) and kNN (Autoencoder-kNN) as initial imputers. For \textit{Reconstruction} and \textit{Prediction} tasks, we used ReLU and Sigmoid as the activation functions, respectively. We trained models for ten epochs using Adam Optimizer and a learning rate of 0.001. We selected the model weights for the epoch where the training loss was the minimum. In rare cases when Autoencoder failed to converge, we resorted to simple kNN for imputation.

\medskip
\begin{algorithm}[H]
    \caption{Participant Level Data Imputation (PLDI)}
    \label{algo:impute}
    \small
    \raggedright
    \SetAlgoLined
    \KwIn{(1) Dataset $D$ with missing values (2) Feature Set $F$ (3) Imputation Strategy $I$}
    \KwOut{Imputed Dataset $\hat{D}$}
     $P \leftarrow \text{ set of participants in } D$ \\
    $D^F \leftarrow D[F]$ \tcp*{Susbset of $D$ with feature set as $F$}
    $\hat{D}^F \leftarrow [ ]$ \tcp*{Empty dataset for storing imputed data}
    
    \For{$p$ \textbf{in} {} $P$}{
        $D^F_p \leftarrow$ data for participant $p$ in $D^F$; \\
        $I.fit(D^F_p)$ \tcp*{Fitting imputation strategy on data}
      
        $\hat{D^F_p } \leftarrow$  $I.transform(D^F_p)$ \tcp*{Apply data imputation strategy $I$}
        $\hat{D^F} \leftarrow \hat{D^F} \cup \hat{D}^F_p$ \tcp*{Concatenate each participant's imputed data}     
    }
    $\hat{D} \leftarrow D$ \tcp*{initializing $\hat{D}$ as $D$}
    $\hat{D}[F] \leftarrow \hat{D}_F$ \tcp*{Updating with imputed values for feature set $F$}
\end{algorithm}
\medskip

We applied our imputation strategies at the participant level (Algorithm \ref{algo:impute}), i.e., for each participant, we only used their own data to fill in missing values within their respective datasets. We decided to conduct imputation at the participant level because each participant can have their own unique set of behavioral habits and routines. Algorithm \ref{algo:reconstruction} and Algorithm \ref{algo:prediction}  show algorithms for reconstruction and prediction tasks, respectively.

It is important to note that while all the imputation strategies increased data availability in datasets, they do not guarantee 100\% data completeness. This is because some users are completely missing a feature; hence, we cannot impute those features following our approach of within-person imputation. One approach could be to perform cross-user imputation for those features, as demonstrated by Rashid~et~al~\cite{rashid2020predicting}. However, given our objective of within-user predictions (equivalent to personalized models), we decided against that approach. This is also representative of a real-world use case, where imputation algorithms might not have data available from other participants in the study. Hence, for a fair direct comparison and to understand the effectiveness of the imputation strategies, we simply decided to pass these imputed datasets to the GLOBEM platform for prediction, and the remaining incompleteness is imputed using the depression modeling algorithms' respective default strategies implemented in the platform.

\begin{algorithm}[h]
    \caption{Reconstruction Objective Algorithm}
    \label{algo:reconstruction}
    \small
    \raggedright
    \SetAlgoLined
    \KwIn{(1) Dataset $D$ (2) Feature Set $F$ (3) Imputation Strategy $I$ (4) Amputation Strategy $A$}
    
    \KwOut{Reconstruction Loss $R$}
    $S = A(D)$ \tcp*{S is indices to be removed decided by amputation strategy $A$}

    $J = D - S$ \tcp*{Removing data at indices S}
    
    $\hat{J} \leftarrow PLDI(J,F,I)$ \tcp*{Algorithm 1}
    $R \leftarrow ReconstructionLoss(J, \hat{J}, S)$ \tcp*{Calculates per participant r-RMSE for indices in S}
\end{algorithm}

\begin{algorithm}[h]
    \caption{Prediction Objective Algorithm}
    \label{algo:prediction}
    \small
    \raggedright
    \SetAlgoLined
    \KwIn{(1) Dataset $D$ with missing values (2) Feature Set $F$ (3) Imputation Strategy $I$ (4) prediction objective $l$}
    
    \KwOut{Predictive Performance Measure $M$}
    
    $\hat{D} \leftarrow PLDI(D,F,I)$ \tcp*{Algorithm 1}
    $M \leftarrow GLOBEM(\hat{D}, l)$ \tcp*{Passing imputed dataset and prediction objective (depression) to GLOBEM Algorithms (e.g., Reorder)}
\end{algorithm}

\subsubsection{GLOBEM's Imputation Strategies:}
In the GLOBEM Platform, imputation happens after the data is divided into input vectors for training models. For the Reorder algorithm, the imputation happens after the training data preparation step, where the participants' data is already divided into 28-day sized input vectors for model training. GLOBEM imputes missing data with the column median across the 28-day input vector, resorting to zeroes if the entire column is empty. On the other hand, the implementation of the Chikersal algorithm uses mean across the entire dataset as an imputation strategy after it has conservatively removed rows and columns having more than 50\% missingness. We refer to the default imputation strategies of Reorder and Chikersal algorithms as the ``GLOBEM-R'' and ``GLOBEM-C'', respectively, and use them as a baseline for their respective comparisons.

For clarity, we want to emphasize that we refer to methods for imputation (e.g., kNN, MICE, etc.) as imputation strategies while referring to methods for depression detection in the GLOBEM platform as algorithms (Chikeral and Reorder).

\section{Results and Evaluation}
\label{sec:results-and-evaluation}

In this section, we present the results of our comparative analysis of different data imputation algorithms, focusing on their impact on data availability, feature-data reconstruction, and prediction performance for the behavioral variable, i.e., depression.

\subsection {Reconstruction Task}
\label{subsubsec:reconstruction-objective}

We first present the relationship between feature data availability and reconstruction RMSE (r-RMSE) for synthetically imputed data per participant. We used one MCAR and two MNAR strategies (MNAR (i) and MNAR (ii)) amputation strategies. We present r-RMSE vs data availability for the Reorder feature set for INS-W\_1 dataset in Figure \ref{fig:reconstruction-dl-feat}, . The trends with other datasets are similar and are included in supplementary materials. 

\begin{figure}[h]
  \centering
  \includegraphics[width=0.8\linewidth]{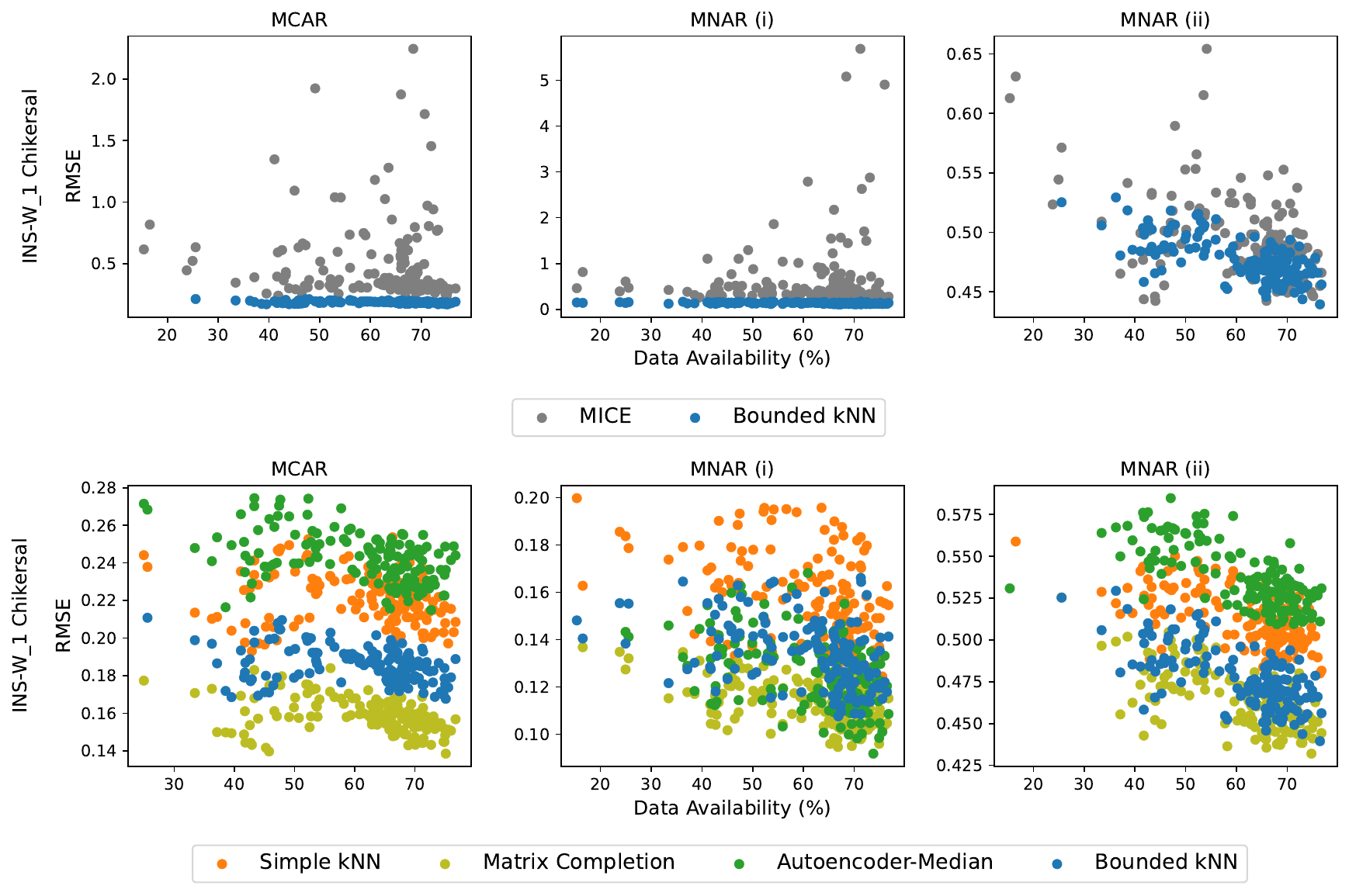}
  \Description{A scatter-plot of INS-W_1 per-user data reconstruction RMSE for different imputation strategies  under MCAR, MNAR (i), and MNAR (ii) amputation. Simple kNN, Matrix Completion, and Autoencoder-Median strategies (bottom) for the GLOBEM Chikersal Algorithm demonstrate significantly better performance when compared to imputation by GLOBEM-R and MICE (top). Also, there is a general inversely proportional relationship between reconstruction and data availability. Different amputation strategies lead to different trends in imputation.}
  \caption{Comparison between imputation strategies for Reconstruction RMSE for Chikersal feature set.}
  \label{fig:reconstruction-ml-chikersal}
\end{figure}

For all three amputation strategies in the Reorder feature set, the GLOBEM-R strategy performed the worst as it replaces feature values with zero if the median of the column is not available in the past 28-day data, most of the time, making feature values out of bounds of the original distribution of the feature. Similarly, MICE performed worse than other strategies as MICE builds regression models, and sometimes, these models predict values out of the bounds of the feature column. For Matrix Completion, Autoencoder-Median, Simple kNN, and Bounded kNN for the Reorder, we observed that the choice of the amputation strategy impacted the performance of the imputation strategies for reconstruction. In the MAR and MNAR (i) amputation strategy, bounded-kNN and Matrix Completion performed the best, whereas, for the MNAR (ii) strategy, all imputation strategies (except MICE and GLOBEM-R) had similar performance. For the Chikersal feature set, this effect is even more prominent (Figure \ref{fig:reconstruction-ml-chikersal})\footnote{We do not show GLOBEM-C in Figure \ref{fig:reconstruction-dl-feat} as the implementation of GLOBEM-C is tied to the data removal strategy of the Chikersal algorithm in the GLOBEM platform. We implemented a proxy to that strategy, and it had worse performance (in a different scale) than all other strategies and thus was not included in the figure.}. For example, Autoencoder-Median performs better for MNAR (i) amputation strategy than Simple KNN but performs worse for MNAR (ii) amputation strategy. The reconstruction graphs for other datasets for Chikersal feature sets are also included in the supplementary materials.
 
In GLOBEM datasets, we checked that the missingness was not completely at random (MCAR); thus, the missing not at random (MNAR) amputation strategies used might be a better way of amputation. Inducing the missingness pattern similar to that found in the dataset using an amputation strategy is inherently challenging and a limitation in this research field. Researchers, however, should be aware that using incorrect amputation strategies can lead to biased and incorrect conclusions.

\begin{figure}[h]
  \centering
  \includegraphics[width=0.8\linewidth]{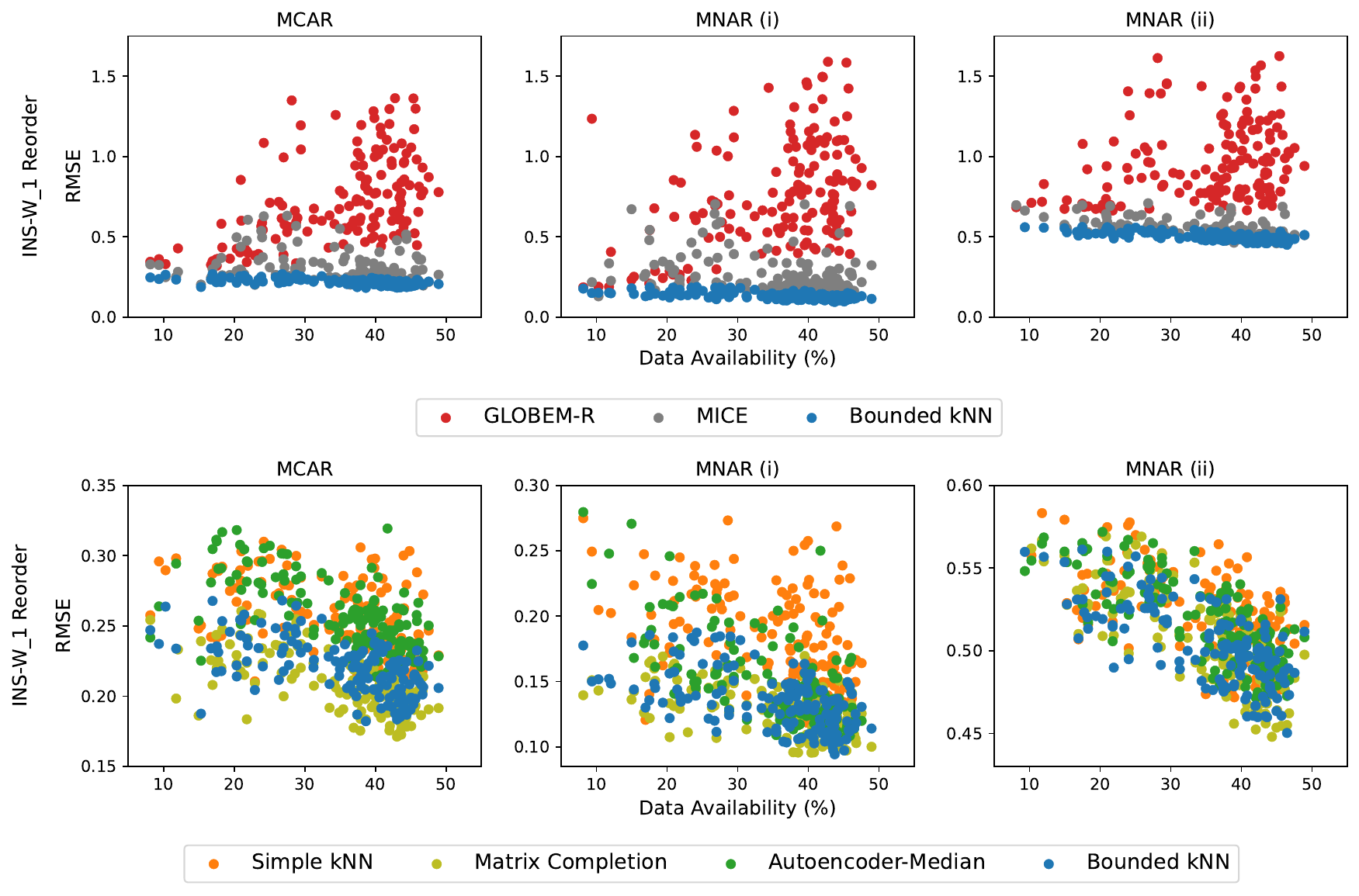}
  \Description{A scatter-plot of INS-W_1 per-user data reconstruction RMSE for different imputation strategies under MCAR, MNAR (i), and MNAR (ii) amputation. Simple kNN, Matrix Completion, and Autoencoder-Median strategies (bottom) for the GLOBEM Reorder Algorithm demonstrate significantly better performance when compared to imputation by GLOBEM-R and MICE (top). Also, there is a general inversely proportional relationship between reconstruction and data availability. Different amputation strategies lead to different trends in imputation. }
  \caption{Comparison between imputation strategies for Reconstruction RMSE for Reorder feature set.}
  \label{fig:reconstruction-dl-feat}
\end{figure}

\begin{figure}[h]
  \centering
  \includegraphics[width=0.8\linewidth]{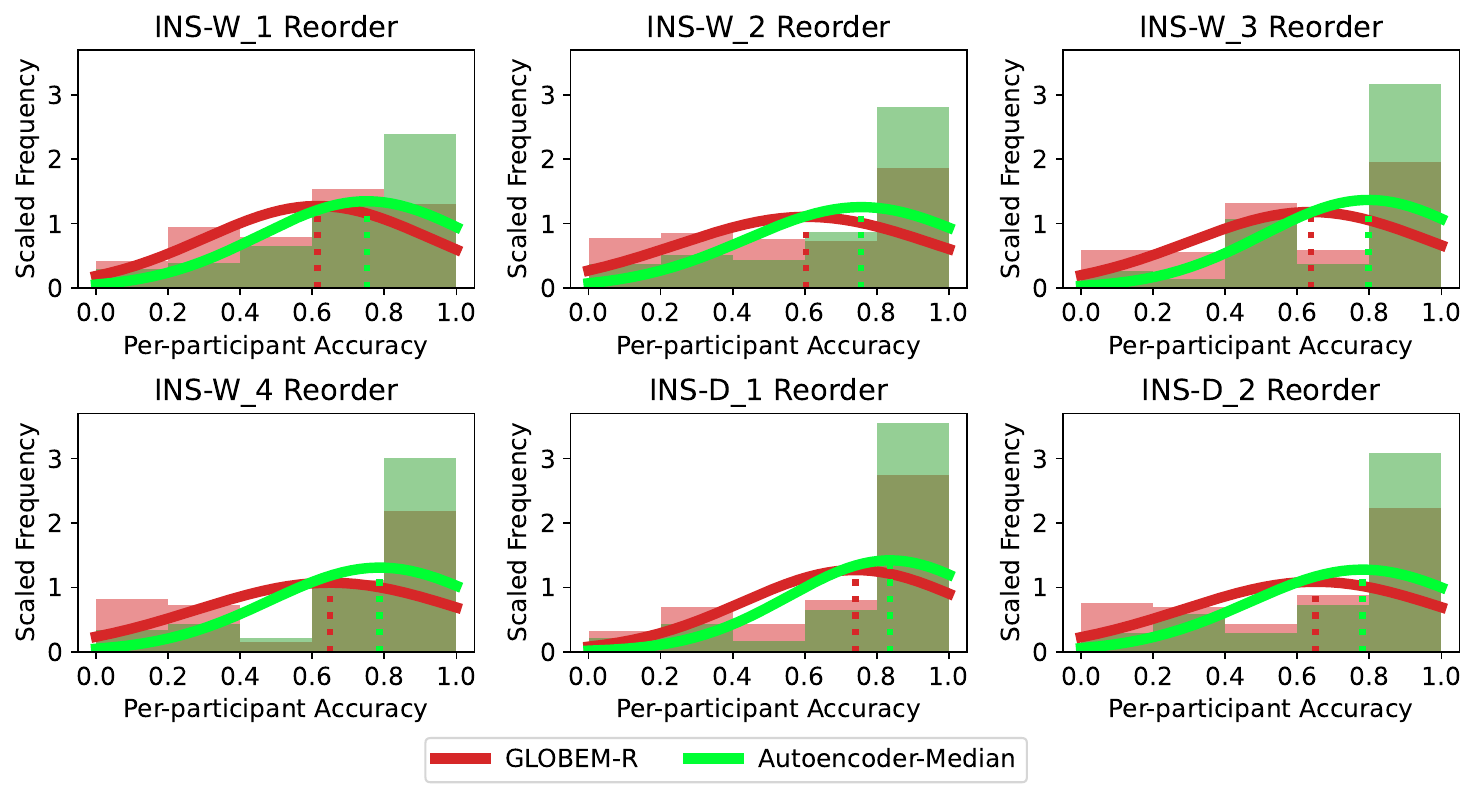}
  \Description{ This figure contains histograms and fitted Gaussian on per-user accuracies when GLOBEM-R and Autoencoder-Median strategies are used. The figure has six subplots, one for each dataset. The best performance improvements when using Autoencoder-Median over GLOBEM-R are clearly evident for all datasets.}
  \caption{Histograms and fitted Gaussian of balanced accuracies per participant in GLOBEM datasets when using Autoencoder-Median and GLOBEM-R as imputation strategies for Reorder algorithm.}
  \label{fig:per-user-accuracy}
\end{figure}

\subsection {Prediction Task}
\label{subsubsec:prediction-objectives}
For the Prediction Task, we considered two scenarios: (1) Depression prediction for future weeks using past weeks' data (outlined as \textit{within-person} task in GLOBEM), and (2) Real-time weekly Depression prediction using past weeks.

\subsubsection{Within-Person Depression Prediction}

We compared the performances of the Reorder and Chikersal models trained on data imputed using different strategies for within-person depression prediction: using the first 80\% of weeks (the past) to predict the last 20\% of weeks (as the future). Tables~\ref{tab:reorder-accs} and Table~\ref{tab:chikersal-accs} compare the success of different imputation strategies in improving the predictive performance of Reorder and Chikersal models, respectively. For Reorder models, Autoencoder-based strategies (Autoencoder-Median and Autoencoder-kNN) consistently showed the best or comparable to best performances across all datasets. Autoencoder-Median imputation increased AUC performance over the baseline approach (GLOBEM-R) used in original work by $25.2\%$, $31.1\%$, $27.1\%$, $22\%$, $19.2\%$, and $17.1\%$ for W\_1, W\_2, W\_3, W\_4, D\_1, and D\_2 datasets respectively. This notable increase in the predictive performance of models emphasizes the importance of choosing the right imputation strategy for datasets. The default GLOBEM-R strategy (median for 28 days features) consistently showed inferior performance compared to all imputation algorithms, including using the median across entire columns (Median). This suggests that relying solely on the median of 28 days for imputation might result in the loss of behavioral information from other days, compromising the accuracy of the imputation process. In Figure \ref{fig:per-user-accuracy}, we demonstrate histograms and fitted Gaussian to compare the distribution of prediction accuracies for individual participants with the Reorder algorithm.  Wilcoxon signed-rank test revealed that the Autoencoder-Median strategy led to a signifciant improvement in performance compared to the baseline GLOBEM-R strategy, for all datasets ($p<0.001$ in all cases and $p=0.009$ in INS\_D\_1). This suggests that Autoencoder-Median imputation better preserves the datasets' behavioral and temporal signals crucial for depression prediction and leads to consistent improvement for most participants. The Reorder algorithm with our Autoencoder-based imputation strategy outperforms all the 19 algorithms -- benchmarked in the original paper and the platform -- for the task of future depression prediction using past data of participants~\cite{xu2023globem}. 

\begin{table}[h]
\caption{Comparison of Balanced Accuracy and AUC when different imputation strategies are used in Reorder algorithm. Avail(\%) denotes the availability of the dataset before passing it to the GLOBEM pipeline and applying the default imputation strategy.}
\resizebox{\linewidth}{!}{
\begin{tabular}{|
>{\columncolor[HTML]{FFFFFF}}c |
>{\columncolor[HTML]{FFFFFF}}c 
>{\columncolor[HTML]{FFFFFF}}c 
>{\columncolor[HTML]{B8CC75}}c 
>{\columncolor[HTML]{FFFFFF}}c 
>{\columncolor[HTML]{FFFFFF}}c 
>{\columncolor[HTML]{B0CA76}}c 
>{\columncolor[HTML]{FFFFFF}}c 
>{\columncolor[HTML]{FFFFFF}}c 
>{\columncolor[HTML]{FFD966}}c 
>{\columncolor[HTML]{FFFFFF}}c 
>{\columncolor[HTML]{FFFFFF}}c 
>{\columncolor[HTML]{DED36D}}c |
>{\columncolor[HTML]{FFFFFF}}c 
>{\columncolor[HTML]{FFFFFF}}c c
>{\columncolor[HTML]{FFFFFF}}c 
>{\columncolor[HTML]{FFFFFF}}c c|}
\hline
\textbf{Reorder}                               & \multicolumn{12}{c|}{\cellcolor[HTML]{FFFFFF}\textbf{INS-W}}                                                                                                                                                                                                                                                                                                                                                                                                                                                                                                                                                                                                                                                                                         & \multicolumn{6}{c|}{\cellcolor[HTML]{FFFFFF}\textbf{INS-D}}                                                                                                                                                                                                                                                                                                    \\ \hline
\textbf{Strategies/Dataset}                     & \multicolumn{3}{c|}{\cellcolor[HTML]{FFFFFF}\textbf{W\_1}}                                                                                                                               & \multicolumn{3}{c|}{\cellcolor[HTML]{FFFFFF}\textbf{W\_2}}                                                                                                                               & \multicolumn{3}{c|}{\cellcolor[HTML]{FFFFFF}\textbf{W\_3}}                                                                                                                               & \multicolumn{3}{c|}{\cellcolor[HTML]{FFFFFF}\textbf{W\_4}}                                                                                                          & \multicolumn{3}{c|}{\cellcolor[HTML]{FFFFFF}\textbf{D\_1}}                                                                                                                               & \multicolumn{3}{c|}{\cellcolor[HTML]{FFFFFF}\textbf{D\_2}}                                                                                                          \\ \hline
\multicolumn{1}{|l|}{\cellcolor[HTML]{FFFFFF}} & \multicolumn{1}{c|}{\cellcolor[HTML]{FFFFFF}\textbf{Acc}} & \multicolumn{1}{c|}{\cellcolor[HTML]{FFFFFF}\textbf{AUC}} & \multicolumn{1}{c|}{\cellcolor[HTML]{FFFFFF}\textbf{Avail (\%)}} & \multicolumn{1}{c|}{\cellcolor[HTML]{FFFFFF}\textbf{Acc}} & \multicolumn{1}{c|}{\cellcolor[HTML]{FFFFFF}\textbf{AUC}} & \multicolumn{1}{c|}{\cellcolor[HTML]{FFFFFF}\textbf{Avail (\%)}} & \multicolumn{1}{c|}{\cellcolor[HTML]{FFFFFF}\textbf{Acc}} & \multicolumn{1}{c|}{\cellcolor[HTML]{FFFFFF}\textbf{AUC}} & \multicolumn{1}{c|}{\cellcolor[HTML]{FFFFFF}\textbf{Avail (\%)}} & \multicolumn{1}{c|}{\cellcolor[HTML]{FFFFFF}\textbf{Acc}} & \multicolumn{1}{c|}{\cellcolor[HTML]{FFFFFF}\textbf{AUC}} & \cellcolor[HTML]{FFFFFF}\textbf{Avail (\%)} & \multicolumn{1}{c|}{\cellcolor[HTML]{FFFFFF}\textbf{Acc}} & \multicolumn{1}{c|}{\cellcolor[HTML]{FFFFFF}\textbf{AUC}} & \multicolumn{1}{c|}{\cellcolor[HTML]{FFFFFF}\textbf{Avail (\%)}} & \multicolumn{1}{c|}{\cellcolor[HTML]{FFFFFF}\textbf{Acc}} & \multicolumn{1}{c|}{\cellcolor[HTML]{FFFFFF}\textbf{AUC}} & \cellcolor[HTML]{FFFFFF}\textbf{Avail (\%)} \\ \hline
\textbf{GLOBEM-R}                              & \multicolumn{1}{c|}{\cellcolor[HTML]{FFFFFF}0.616}        & \multicolumn{1}{c|}{\cellcolor[HTML]{FFFFFF}0.657}        & \multicolumn{1}{c|}{\cellcolor[HTML]{E16C66}33.71}               & \multicolumn{1}{c|}{\cellcolor[HTML]{FFFFFF}0.606}        & \multicolumn{1}{c|}{\cellcolor[HTML]{FFFFFF}0.619}        & \multicolumn{1}{c|}{\cellcolor[HTML]{E57A66}37.82}               & \multicolumn{1}{c|}{\cellcolor[HTML]{FFFFFF}0.639}        & \multicolumn{1}{c|}{\cellcolor[HTML]{FFFFFF}0.671}        & \multicolumn{1}{c|}{\cellcolor[HTML]{E06666}31.77}               & \multicolumn{1}{c|}{\cellcolor[HTML]{FFFFFF}0.644}        & \multicolumn{1}{c|}{\cellcolor[HTML]{FFFFFF}0.692}        & \cellcolor[HTML]{E37266}35.43               & \multicolumn{1}{c|}{\cellcolor[HTML]{FFFFFF}0.732}        & \multicolumn{1}{c|}{\cellcolor[HTML]{FFFFFF}0.759}        & \multicolumn{1}{c|}{\cellcolor[HTML]{E57B66}38.11}               & \multicolumn{1}{c|}{\cellcolor[HTML]{FFFFFF}0.641}        & \multicolumn{1}{c|}{\cellcolor[HTML]{FFFFFF}0.702}        & \cellcolor[HTML]{E78366}40.24              \\ \hline
\textbf{Median}                                & \multicolumn{1}{c|}{\cellcolor[HTML]{FFFFFF}0.674}        & \multicolumn{1}{c|}{\cellcolor[HTML]{FFFFFF}0.768}        & \multicolumn{1}{c|}{\cellcolor[HTML]{B8CC75}87.42}               & \multicolumn{1}{c|}{\cellcolor[HTML]{FFFFFF}0.676}        & \multicolumn{1}{c|}{\cellcolor[HTML]{FFFFFF}0.765}        & \multicolumn{1}{c|}{\cellcolor[HTML]{B0CA76}89.98}               & \multicolumn{1}{c|}{\cellcolor[HTML]{FFFFFF}0.715}        & \multicolumn{1}{c|}{\cellcolor[HTML]{FFFFFF}0.763}        & \multicolumn{1}{c|}{\cellcolor[HTML]{FFD966}65.41}               & \multicolumn{1}{c|}{\cellcolor[HTML]{FFFFFF}0.751}        & \multicolumn{1}{c|}{\cellcolor[HTML]{FFFFFF}0.811}        & 75.54                                       & \multicolumn{1}{c|}{\cellcolor[HTML]{FFFFFF}0.768}        & \multicolumn{1}{c|}{\cellcolor[HTML]{FFFFFF}0.868}        & \multicolumn{1}{c|}{\cellcolor[HTML]{D4D16F}78.61}               & \multicolumn{1}{c|}{\cellcolor[HTML]{FFFFFF}0.722}        & \multicolumn{1}{c|}{\cellcolor[HTML]{FFFFFF}0.802}        & \cellcolor[HTML]{D7D26E}77.77               \\ \hline
\textbf{Simple kNN}                           & \multicolumn{1}{c|}{\cellcolor[HTML]{FFFFFF}0.723}        & \multicolumn{1}{c|}{\cellcolor[HTML]{FFFFFF}0.798}        & \multicolumn{1}{c|}{\cellcolor[HTML]{B8CC75}87.42}               & \multicolumn{1}{c|}{\cellcolor[HTML]{FFFFFF}0.708}        & \multicolumn{1}{c|}{\cellcolor[HTML]{FFFFFF}0.767}        & \multicolumn{1}{c|}{\cellcolor[HTML]{B0CA76}89.98}               & \multicolumn{1}{c|}{\cellcolor[HTML]{FFFFFF}0.745}        & \multicolumn{1}{c|}{\cellcolor[HTML]{FFFFFF}0.792}        & \multicolumn{1}{c|}{\cellcolor[HTML]{FFD966}65.41}               & \multicolumn{1}{c|}{\cellcolor[HTML]{FFFFFF}0.744}        & \multicolumn{1}{c|}{\cellcolor[HTML]{FFFFFF}0.813}        & 75.54                                       & \multicolumn{1}{c|}{\cellcolor[HTML]{FFFFFF}0.788}        & \multicolumn{1}{c|}{\cellcolor[HTML]{FFFFFF}0.873}        & \multicolumn{1}{c|}{\cellcolor[HTML]{D4D16F}78.61}               & \multicolumn{1}{c|}{\cellcolor[HTML]{FFFFFF}0.757}        & \multicolumn{1}{c|}{\cellcolor[HTML]{FFFFFF}0.834}        & \cellcolor[HTML]{D7D26E}77.77               \\ \hline
\textbf{Bounded kNN}                           & \multicolumn{1}{c|}{\cellcolor[HTML]{FFFFFF}0.697}        & \multicolumn{1}{c|}{\cellcolor[HTML]{FFFFFF}0.769}        & \multicolumn{1}{c|}{\cellcolor[HTML]{FDD366}63.68}               & \multicolumn{1}{c|}{\cellcolor[HTML]{FFFFFF}0.693}        & \multicolumn{1}{c|}{\cellcolor[HTML]{FFFFFF}0.745}        & \multicolumn{1}{c|}{\cellcolor[HTML]{FBD966}66.56}               & \multicolumn{1}{c|}{\cellcolor[HTML]{FFFFFF}0.680}        & \multicolumn{1}{c|}{\cellcolor[HTML]{FFFFFF}0.726}        & \multicolumn{1}{c|}{\cellcolor[HTML]{ED9966}46.83}               & \multicolumn{1}{c|}{\cellcolor[HTML]{FFFFFF}\textbf{0.818}}        & \multicolumn{1}{c|}{\cellcolor[HTML]{FFFFFF}0.857}        & \cellcolor[HTML]{F5B666}55.34               & \multicolumn{1}{c|}{\cellcolor[HTML]{FFFFFF}0.804}        & \multicolumn{1}{c|}{\cellcolor[HTML]{FFFFFF}0.876}        & \multicolumn{1}{c|}{\cellcolor[HTML]{F9C366}58.85}               & \multicolumn{1}{c|}{\cellcolor[HTML]{FFFFFF}0.700}        & \multicolumn{1}{c|}{\cellcolor[HTML]{FFFFFF}0.758}        & \cellcolor[HTML]{F9C566}59.54               \\ \hline
\textbf{MICE}                                  & \multicolumn{1}{c|}{\cellcolor[HTML]{FFFFFF}0.724}        & \multicolumn{1}{c|}{\cellcolor[HTML]{FFFFFF}0.794}        & \multicolumn{1}{c|}{\cellcolor[HTML]{B8CC75}87.42}               & \multicolumn{1}{c|}{\cellcolor[HTML]{FFFFFF}0.715}        & \multicolumn{1}{c|}{\cellcolor[HTML]{FFFFFF}0.779}        & \multicolumn{1}{c|}{\cellcolor[HTML]{B0CA76}89.98}               & \multicolumn{1}{c|}{\cellcolor[HTML]{FFFFFF}0.717}        & \multicolumn{1}{c|}{\cellcolor[HTML]{FFFFFF}0.776}        & \multicolumn{1}{c|}{\cellcolor[HTML]{FFD966}65.41}               & \multicolumn{1}{c|}{\cellcolor[HTML]{FFFFFF}0.771}        & \multicolumn{1}{c|}{\cellcolor[HTML]{FFFFFF}0.836}        & 75.54                                       & \multicolumn{1}{c|}{\cellcolor[HTML]{FFFFFF}0.763}        & \multicolumn{1}{c|}{\cellcolor[HTML]{FFFFFF}0.841}        & \multicolumn{1}{c|}{\cellcolor[HTML]{D4D16F}78.61}               & \multicolumn{1}{c|}{\cellcolor[HTML]{FFFFFF}0.770}        & \multicolumn{1}{c|}{\cellcolor[HTML]{FFFFFF}\textbf{0.840}}        & \cellcolor[HTML]{D7D26E}77.77               \\ \hline
\textbf{Matrix Completion}                     & \multicolumn{1}{c|}{\cellcolor[HTML]{FFFFFF}0.703}        & \multicolumn{1}{c|}{\cellcolor[HTML]{FFFFFF}0.782}        & \multicolumn{1}{c|}{\cellcolor[HTML]{B8CC75}87.42}               & \multicolumn{1}{c|}{\cellcolor[HTML]{FFFFFF}0.699}        & \multicolumn{1}{c|}{\cellcolor[HTML]{FFFFFF}0.761}        & \multicolumn{1}{c|}{\cellcolor[HTML]{B0CA76}89.98}               & \multicolumn{1}{c|}{\cellcolor[HTML]{FFFFFF}0.692}        & \multicolumn{1}{c|}{\cellcolor[HTML]{FFFFFF}0.771}        & \multicolumn{1}{c|}{\cellcolor[HTML]{FFD966}65.41}               & \multicolumn{1}{c|}{\cellcolor[HTML]{FFFFFF}0.730}        & \multicolumn{1}{c|}{\cellcolor[HTML]{FFFFFF}0.809}        & 75.54                                       & \multicolumn{1}{c|}{\cellcolor[HTML]{FFFFFF}0.784}        & \multicolumn{1}{c|}{\cellcolor[HTML]{FFFFFF}0.845}        & \multicolumn{1}{c|}{\cellcolor[HTML]{F4D768}68.79}               & \multicolumn{1}{c|}{\cellcolor[HTML]{FFFFFF}0.695}        & \multicolumn{1}{c|}{\cellcolor[HTML]{FFFFFF}0.789}        & \cellcolor[HTML]{F6D867}68.13               \\ \hline
\textbf{Autoencoder-Median}                    & \multicolumn{1}{c|}{\cellcolor[HTML]{FFFFFF}\textbf{0.750}}        & \multicolumn{1}{c|}{\cellcolor[HTML]{FFFFFF}\textbf{0.823}}        & \multicolumn{1}{c|}{\cellcolor[HTML]{B8CC75}87.42}               & \multicolumn{1}{c|}{\cellcolor[HTML]{FFFFFF}\textbf{0.753}}        & \multicolumn{1}{c|}{\cellcolor[HTML]{FFFFFF}\textbf{0.812}}        & \multicolumn{1}{c|}{\cellcolor[HTML]{B0CA76}89.98}               & \multicolumn{1}{c|}{\cellcolor[HTML]{FFFFFF}\textbf{0.807}}        & \multicolumn{1}{c|}{\cellcolor[HTML]{FFFFFF}\textbf{0.853}}        & \multicolumn{1}{c|}{\cellcolor[HTML]{FFD966}65.41}               & \multicolumn{1}{c|}{\cellcolor[HTML]{FFFFFF}0.795}        & \multicolumn{1}{c|}{\cellcolor[HTML]{FFFFFF}0.844}        & 75.54                                       & \multicolumn{1}{c|}{\cellcolor[HTML]{FFFFFF}\textbf{0.841}}        & \multicolumn{1}{c|}{\cellcolor[HTML]{FFFFFF}\textbf{0.906}}        & \multicolumn{1}{c|}{\cellcolor[HTML]{B3CB76}88.99}               & \multicolumn{1}{c|}{\cellcolor[HTML]{FFFFFF}0.764}        & \multicolumn{1}{c|}{\cellcolor[HTML]{FFFFFF}0.822}        & \cellcolor[HTML]{D7D26E}77.77               \\ \hline
\textbf{Autoencoder-kNN}                       & \multicolumn{1}{c|}{\cellcolor[HTML]{FFFFFF}0.733}        & \multicolumn{1}{c|}{\cellcolor[HTML]{FFFFFF}0.810}        & \multicolumn{1}{c|}{\cellcolor[HTML]{B8CC75}87.42}               & \multicolumn{1}{c|}{\cellcolor[HTML]{FFFFFF}0.740}        & \multicolumn{1}{c|}{\cellcolor[HTML]{FFFFFF}0.807}        & \multicolumn{1}{c|}{\cellcolor[HTML]{B0CA76}89.98}               & \multicolumn{1}{c|}{\cellcolor[HTML]{FFFFFF}0.734}        & \multicolumn{1}{c|}{\cellcolor[HTML]{FFFFFF}0.814}        & \multicolumn{1}{c|}{\cellcolor[HTML]{FFD966}65.41}               & \multicolumn{1}{c|}{\cellcolor[HTML]{FFFFFF}0.793}        & \multicolumn{1}{c|}{\cellcolor[HTML]{FFFFFF}\textbf{0.864}}        & 75.54                                       & \multicolumn{1}{c|}{\cellcolor[HTML]{FFFFFF}0.804}        & \multicolumn{1}{c|}{\cellcolor[HTML]{FFFFFF}0.869}        & \multicolumn{1}{c|}{\cellcolor[HTML]{D4D16F}78.61}               & \multicolumn{1}{c|}{\cellcolor[HTML]{FFFFFF}\textbf{0.773}}        & \multicolumn{1}{c|}{\cellcolor[HTML]{FFFFFF}0.839}        & \cellcolor[HTML]{D7D26E}77.77               \\ \hline
\end{tabular}
\Description{This table contains a comparison of different imputation strategies for the "within-person" depression prediction task with Reorder Algorithm. It also has the data availability after imputation for the strategies. We see that Autoencoder-Median and Autoencoder-kNN perform best or comparable to best in most cases}
}
%

\label{tab:reorder-accs}
\end{table}

For the Chikersal algorithm, since datasets already had high data availability, there was limited scope for data imputation. Additionally, we found that the Chikersal algorithm implementation removes columns and rows with less than 50\% data, and thus, even after imputation, the rows might not be used in model training if they are still below the threshold. Unlike Reorder, which models temporal behavior using the last 28 days of feature values, the Chikersal model only uses feature values on the day of the week when depression labels were collected. Due to these reasons, even after we impute the datasets, the imputed datasets are not fully used by the algorithm, leading to less distinctive prediction performances across all imputation strategies. Autoencoder-based strategies and Matrix Completion have slightly higher performance compared to other algorithms for most datasets. The Reorder algorithm, however, which has considerably high missing data, less conservative filtering, and models temporal behavioral changes, benefits the most from using different imputation strategies compared to the Chikersal algorithm. Our findings not only indicate that our proposed Autoencoder-based approach is a viable imputation approach leading to significant performance improvement, but also highlight that imputation strategies are not a one-size-fits-all solution and should be a core part of the model building and training process. 

In our evaluations, we imputed the data before passing it to the GLOBEM pipeline, where the platform divides the dataset into training and test data. Since the imputation considered all the data for each person, we wanted to investigate whether imputation strategies were causing data leakage, thus leading to a boost in the model performances. To evaluate potential data leakage, we fitted our imputation strategies on only the training data and then used it to impute both the train and test data. We present our findings on the performance of the Reorder algorithm in Table~\ref{tab:reorder-data-leakage} and for the Chikersal algorithm in supplementary materials. We observed that the model performances, in this case, were comparable with the model performances when the entire dataset was used for imputation. This confirms that the boost in performance is not due to data leakage but rather due to better imputation strategies.

\begin{table}[h]
\caption{Comparison of Balanced Accuracy and AUC when different imputation strategies are used in Chikersal algorithm. Avail(\%) denotes the dataset availability before passing it to the GLOBEM pipeline and applying the default imputation strategy.}
\resizebox{\linewidth}{!}{
\begin{tabular}{|c|cccccccccccc|cccccc|}
\hline
\rowcolor[HTML]{FFFFFF} 
\textbf{Chikersal}                             & \multicolumn{12}{c|}{\cellcolor[HTML]{FFFFFF}\textbf{INS-W}}                                                                                                                                                                                                                                                                                                                                                                                                                                                                                                                                                                                                                                                                           & \multicolumn{6}{c|}{\cellcolor[HTML]{FFFFFF}\textbf{INS-D}}                                                                                                                                                                                                                                                                                      \\ \hline
\rowcolor[HTML]{FFFFFF} 
\textbf{Strategies/Dataset}                    & \multicolumn{3}{c|}{\cellcolor[HTML]{FFFFFF}\textbf{W\_1}}                                                                                                                               & \multicolumn{3}{c|}{\cellcolor[HTML]{FFFFFF}\textbf{W\_2}}                                                                                                                               & \multicolumn{3}{c|}{\cellcolor[HTML]{FFFFFF}\textbf{W\_3}}                                                                                                                               & \multicolumn{3}{c|}{\cellcolor[HTML]{FFFFFF}\textbf{W\_4}}                                                                                            & \multicolumn{3}{c|}{\cellcolor[HTML]{FFFFFF}\textbf{D\_1}}                                                                                                                               & \multicolumn{3}{c|}{\cellcolor[HTML]{FFFFFF}\textbf{D\_2}}                                                                                            \\ \hline
\rowcolor[HTML]{FFFFFF} 
\multicolumn{1}{|l|}{\cellcolor[HTML]{FFFFFF}} & \multicolumn{1}{c|}{\cellcolor[HTML]{FFFFFF}\textbf{Acc}} & \multicolumn{1}{c|}{\cellcolor[HTML]{FFFFFF}\textbf{AUC}} & \multicolumn{1}{c|}{\cellcolor[HTML]{FFFFFF}\textbf{Avail (\%)}} & \multicolumn{1}{c|}{\cellcolor[HTML]{FFFFFF}\textbf{Acc}} & \multicolumn{1}{c|}{\cellcolor[HTML]{FFFFFF}\textbf{AUC}} & \multicolumn{1}{c|}{\cellcolor[HTML]{FFFFFF}\textbf{Avail (\%)}} & \multicolumn{1}{c|}{\cellcolor[HTML]{FFFFFF}\textbf{Acc}} & \multicolumn{1}{c|}{\cellcolor[HTML]{FFFFFF}\textbf{AUC}} & \multicolumn{1}{c|}{\cellcolor[HTML]{FFFFFF}\textbf{Avail (\%)}} & \multicolumn{1}{c|}{\cellcolor[HTML]{FFFFFF}\textbf{Acc}} & \multicolumn{1}{c|}{\cellcolor[HTML]{FFFFFF}\textbf{AUC}} & \textbf{Avail (\%)}           & \multicolumn{1}{c|}{\cellcolor[HTML]{FFFFFF}\textbf{Acc}} & \multicolumn{1}{c|}{\cellcolor[HTML]{FFFFFF}\textbf{AUC}} & \multicolumn{1}{c|}{\cellcolor[HTML]{FFFFFF}\textbf{Avail (\%)}} & \multicolumn{1}{c|}{\cellcolor[HTML]{FFFFFF}\textbf{Acc}} & \multicolumn{1}{c|}{\cellcolor[HTML]{FFFFFF}\textbf{AUC}} & \textbf{Avail (\%)}           \\ \hline
\rowcolor[HTML]{FFFFFF} 
\textbf{GLOBEM-C}                              & \multicolumn{1}{c|}{\cellcolor[HTML]{FFFFFF}0.656}        & \multicolumn{1}{c|}{\cellcolor[HTML]{FFFFFF}0.727}        & \multicolumn{1}{c|}{\cellcolor[HTML]{FAC766}60.17}               & \multicolumn{1}{c|}{\cellcolor[HTML]{FFFFFF}0.611}        & \multicolumn{1}{c|}{\cellcolor[HTML]{FFFFFF}0.680}        & \multicolumn{1}{c|}{\cellcolor[HTML]{FFD966}65.51}               & \multicolumn{1}{c|}{\cellcolor[HTML]{FFFFFF}0.641}        & \multicolumn{1}{c|}{\cellcolor[HTML]{FFFFFF}0.696}        & \multicolumn{1}{c|}{\cellcolor[HTML]{FBCD66}61.79}               & \multicolumn{1}{c|}{\cellcolor[HTML]{FFFFFF}0.690}        & \multicolumn{1}{c|}{\cellcolor[HTML]{FFFFFF}0.763}        & \cellcolor[HTML]{FFD966}65.53 & \multicolumn{1}{c|}{\cellcolor[HTML]{FFFFFF}0.649}        & \multicolumn{1}{c|}{\cellcolor[HTML]{FFFFFF}0.696}        & \multicolumn{1}{c|}{\cellcolor[HTML]{EBD56A}71.63}               & \multicolumn{1}{c|}{\cellcolor[HTML]{FFFFFF}0.651}        & \multicolumn{1}{c|}{\cellcolor[HTML]{FFFFFF}0.726}        & \cellcolor[HTML]{DFD36C}75.32 \\ \hline
\rowcolor[HTML]{FFFFFF} 
\textbf{Median}                                & \multicolumn{1}{c|}{\cellcolor[HTML]{FFFFFF}0.630}        & \multicolumn{1}{c|}{\cellcolor[HTML]{FFFFFF}0.701}        & \multicolumn{1}{c|}{\cellcolor[HTML]{A2C779}94.18}               & \multicolumn{1}{c|}{\cellcolor[HTML]{FFFFFF}\textbf{0.671}}        & \multicolumn{1}{c|}{\cellcolor[HTML]{FFFFFF}0.728}        & \multicolumn{1}{c|}{\cellcolor[HTML]{98C57C}97.36}               & \multicolumn{1}{c|}{\cellcolor[HTML]{FFFFFF}0.655}        & \multicolumn{1}{c|}{\cellcolor[HTML]{FFFFFF}0.697}        & \multicolumn{1}{c|}{\cellcolor[HTML]{97C57C}97.55}               & \multicolumn{1}{c|}{\cellcolor[HTML]{FFFFFF}0.705}        & \multicolumn{1}{c|}{\cellcolor[HTML]{FFFFFF}0.799}        & \cellcolor[HTML]{98C57B}97.19 & \multicolumn{1}{c|}{\cellcolor[HTML]{FFFFFF}0.644}        & \multicolumn{1}{c|}{\cellcolor[HTML]{FFFFFF}0.747}        & \multicolumn{1}{c|}{\cellcolor[HTML]{97C57C}97.70}               & \multicolumn{1}{c|}{\cellcolor[HTML]{FFFFFF}\textbf{0.716}}        & \multicolumn{1}{c|}{\cellcolor[HTML]{FFFFFF}\textbf{0.785}}        & \cellcolor[HTML]{95C57C}98.16 \\ \hline
\rowcolor[HTML]{FFFFFF} 
\textbf{Simple kNN}                           & \multicolumn{1}{c|}{\cellcolor[HTML]{FFFFFF}0.640}        & \multicolumn{1}{c|}{\cellcolor[HTML]{FFFFFF}0.698}        & \multicolumn{1}{c|}{\cellcolor[HTML]{A2C779}94.18}               & \multicolumn{1}{c|}{\cellcolor[HTML]{FFFFFF}0.630}        & \multicolumn{1}{c|}{\cellcolor[HTML]{FFFFFF}0.666}        & \multicolumn{1}{c|}{\cellcolor[HTML]{98C57C}97.36}               & \multicolumn{1}{c|}{\cellcolor[HTML]{FFFFFF}0.643}        & \multicolumn{1}{c|}{\cellcolor[HTML]{FFFFFF}0.703}        & \multicolumn{1}{c|}{\cellcolor[HTML]{97C57C}97.55}               & \multicolumn{1}{c|}{\cellcolor[HTML]{FFFFFF}0.694}        & \multicolumn{1}{c|}{\cellcolor[HTML]{FFFFFF}0.792}        & \cellcolor[HTML]{98C57B}97.19 & \multicolumn{1}{c|}{\cellcolor[HTML]{FFFFFF}0.649}        & \multicolumn{1}{c|}{\cellcolor[HTML]{FFFFFF}0.722}        & \multicolumn{1}{c|}{\cellcolor[HTML]{93C47D}98.72}               & \multicolumn{1}{c|}{\cellcolor[HTML]{FFFFFF}0.608}        & \multicolumn{1}{c|}{\cellcolor[HTML]{FFFFFF}0.637}        & \cellcolor[HTML]{95C57C}98.33 \\ \hline
\rowcolor[HTML]{FFFFFF} 
\textbf{Bounded kNN}                           & \multicolumn{1}{c|}{\cellcolor[HTML]{FFFFFF}0.679}        & \multicolumn{1}{c|}{\cellcolor[HTML]{FFFFFF}0.753}        & \multicolumn{1}{c|}{\cellcolor[HTML]{B2CA76}89.26}               & \multicolumn{1}{c|}{\cellcolor[HTML]{FFFFFF}0.641}        & \multicolumn{1}{c|}{\cellcolor[HTML]{FFFFFF}0.704}        & \multicolumn{1}{c|}{\cellcolor[HTML]{A9C978}92.18}               & \multicolumn{1}{c|}{\cellcolor[HTML]{FFFFFF}0.670}        & \multicolumn{1}{c|}{\cellcolor[HTML]{FFFFFF}0.713}        & \multicolumn{1}{c|}{\cellcolor[HTML]{B0CA77}89.99}               & \multicolumn{1}{c|}{\cellcolor[HTML]{FFFFFF}0.691}        & \multicolumn{1}{c|}{\cellcolor[HTML]{FFFFFF}0.769}        & \cellcolor[HTML]{ACC977}91.19 & \multicolumn{1}{c|}{\cellcolor[HTML]{FFFFFF}0.653}        & \multicolumn{1}{c|}{\cellcolor[HTML]{FFFFFF}0.691}        & \multicolumn{1}{c|}{\cellcolor[HTML]{A5C879}93.32}               & \multicolumn{1}{c|}{\cellcolor[HTML]{FFFFFF}0.663}        & \multicolumn{1}{c|}{\cellcolor[HTML]{FFFFFF}0.729}        & \cellcolor[HTML]{A1C77A}94.56 \\ \hline
\rowcolor[HTML]{FFFFFF} 
\textbf{MICE}                                  & \multicolumn{1}{c|}{\cellcolor[HTML]{FFFFFF}0.656}        & \multicolumn{1}{c|}{\cellcolor[HTML]{FFFFFF}0.703}        & \multicolumn{1}{c|}{\cellcolor[HTML]{A2C779}94.18}               & \multicolumn{1}{c|}{\cellcolor[HTML]{FFFFFF}0.622}        & \multicolumn{1}{c|}{\cellcolor[HTML]{FFFFFF}0.667}        & \multicolumn{1}{c|}{\cellcolor[HTML]{98C57C}97.36}               & \multicolumn{1}{c|}{\cellcolor[HTML]{FFFFFF}0.656}        & \multicolumn{1}{c|}{\cellcolor[HTML]{FFFFFF}0.698}        & \multicolumn{1}{c|}{\cellcolor[HTML]{97C57C}97.55}               & \multicolumn{1}{c|}{\cellcolor[HTML]{FFFFFF}0.645}        & \multicolumn{1}{c|}{\cellcolor[HTML]{FFFFFF}0.734}        & \cellcolor[HTML]{98C57B}97.19 & \multicolumn{1}{c|}{\cellcolor[HTML]{FFFFFF}0.701}        & \multicolumn{1}{c|}{\cellcolor[HTML]{FFFFFF}0.753}        & \multicolumn{1}{c|}{\cellcolor[HTML]{97C57C}97.70}               & \multicolumn{1}{c|}{\cellcolor[HTML]{FFFFFF}0.584}        & \multicolumn{1}{c|}{\cellcolor[HTML]{FFFFFF}0.638}        & \cellcolor[HTML]{95C57C}98.33 \\ \hline
\rowcolor[HTML]{FFFFFF} 
\textbf{Matrix Completion}                     & \multicolumn{1}{c|}{\cellcolor[HTML]{FFFFFF}\textbf{0.717}}        & \multicolumn{1}{c|}{\cellcolor[HTML]{FFFFFF}\textbf{0.768}}        & \multicolumn{1}{c|}{\cellcolor[HTML]{AFCA77}90.07}               & \multicolumn{1}{c|}{\cellcolor[HTML]{FFFFFF}0.663}        & \multicolumn{1}{c|}{\cellcolor[HTML]{FFFFFF}\textbf{0.740}}        & \multicolumn{1}{c|}{\cellcolor[HTML]{A4C879}93.66}               & \multicolumn{1}{c|}{\cellcolor[HTML]{FFFFFF}\textbf{0.695}}        & \multicolumn{1}{c|}{\cellcolor[HTML]{FFFFFF}\textbf{0.744}}        & \multicolumn{1}{c|}{\cellcolor[HTML]{B1CA76}89.48}               & \multicolumn{1}{c|}{\cellcolor[HTML]{FFFFFF}0.710}        & \multicolumn{1}{c|}{\cellcolor[HTML]{FFFFFF}0.803}        & \cellcolor[HTML]{AAC978}91.77 & \multicolumn{1}{c|}{\cellcolor[HTML]{FFFFFF}0.592}        & \multicolumn{1}{c|}{\cellcolor[HTML]{FFFFFF}0.638}        & \multicolumn{1}{c|}{\cellcolor[HTML]{B0CA77}90.03}               & \multicolumn{1}{c|}{\cellcolor[HTML]{FFFFFF}0.667}        & \multicolumn{1}{c|}{\cellcolor[HTML]{FFFFFF}0.741}        & \cellcolor[HTML]{AFCA77}90.18 \\ \hline
\rowcolor[HTML]{FFFFFF} 
\textbf{Autoencoder-Median}                    & \multicolumn{1}{c|}{\cellcolor[HTML]{FFFFFF}0.703}        & \multicolumn{1}{c|}{\cellcolor[HTML]{FFFFFF}0.754}        & \multicolumn{1}{c|}{\cellcolor[HTML]{A2C779}94.18}               & \multicolumn{1}{c|}{\cellcolor[HTML]{FFFFFF}0.648}        & \multicolumn{1}{c|}{\cellcolor[HTML]{FFFFFF}0.695}        & \multicolumn{1}{c|}{\cellcolor[HTML]{98C57C}97.36}               & \multicolumn{1}{c|}{\cellcolor[HTML]{FFFFFF}0.629}        & \multicolumn{1}{c|}{\cellcolor[HTML]{FFFFFF}0.679}        & \multicolumn{1}{c|}{\cellcolor[HTML]{97C57C}97.55}               & \multicolumn{1}{c|}{\cellcolor[HTML]{FFFFFF}0.686}        & \multicolumn{1}{c|}{\cellcolor[HTML]{FFFFFF}0.762}        & \cellcolor[HTML]{98C57B}97.19 & \multicolumn{1}{c|}{\cellcolor[HTML]{FFFFFF}\textbf{0.742}}        & \multicolumn{1}{c|}{\cellcolor[HTML]{FFFFFF}\textbf{0.821}}        & \multicolumn{1}{c|}{\cellcolor[HTML]{93C47D}98.72}               & \multicolumn{1}{c|}{\cellcolor[HTML]{FFFFFF}0.658}        & \multicolumn{1}{c|}{\cellcolor[HTML]{FFFFFF}0.716}        & \cellcolor[HTML]{95C57C}98.33 \\ \hline
\rowcolor[HTML]{FFFFFF} 
\textbf{Autoencoder-kNN}                       & \multicolumn{1}{c|}{\cellcolor[HTML]{FFFFFF}0.649}        & \multicolumn{1}{c|}{\cellcolor[HTML]{FFFFFF}0.732}        & \multicolumn{1}{c|}{\cellcolor[HTML]{A2C779}94.18}               & \multicolumn{1}{c|}{\cellcolor[HTML]{FFFFFF}0.616}        & \multicolumn{1}{c|}{\cellcolor[HTML]{FFFFFF}0.658}        & \multicolumn{1}{c|}{\cellcolor[HTML]{98C57C}97.36}               & \multicolumn{1}{c|}{\cellcolor[HTML]{FFFFFF}0.692}        & \multicolumn{1}{c|}{\cellcolor[HTML]{FFFFFF}0.738}        & \multicolumn{1}{c|}{\cellcolor[HTML]{97C57C}97.55}               & \multicolumn{1}{c|}{\cellcolor[HTML]{FFFFFF}\textbf{0.745}}        & \multicolumn{1}{c|}{\cellcolor[HTML]{FFFFFF}\textbf{0.810}}        & \cellcolor[HTML]{98C57B}97.19 & \multicolumn{1}{c|}{\cellcolor[HTML]{FFFFFF}0.660}        & \multicolumn{1}{c|}{\cellcolor[HTML]{FFFFFF}0.739}        & \multicolumn{1}{c|}{\cellcolor[HTML]{93C47D}98.72}               & \multicolumn{1}{c|}{\cellcolor[HTML]{FFFFFF}0.559}        & \multicolumn{1}{c|}{\cellcolor[HTML]{FFFFFF}0.617}        & \cellcolor[HTML]{95C57C}98.33 \\ \hline
\end{tabular}
\Description{This table contains a comparison of different imputation strategies for the "within-person" depression prediction task with Chikersal algorithm. It also has the data availability after imputation for the strategies. We see that Autoencoder-Median and Autoencoder-kNN has good performance in most cases but the improvements are not as significant as in Reorder algorithm.}
}

\label{tab:chikersal-accs}

\end{table}

\begin{table*}[]
\caption{Data leakage analysis of Reorder algorithm when imputation is only performed using training data}
\resizebox{\linewidth}{!}{
\begin{tabular}{|
>{\columncolor[HTML]{FFFFFF}}c |
>{\columncolor[HTML]{FFFFFF}}c 
>{\columncolor[HTML]{FFFFFF}}c 
>{\columncolor[HTML]{FFFFFF}}c 
>{\columncolor[HTML]{FFFFFF}}c 
>{\columncolor[HTML]{FFFFFF}}c 
>{\columncolor[HTML]{FFFFFF}}c 
>{\columncolor[HTML]{FFFFFF}}c 
>{\columncolor[HTML]{FFFFFF}}c |
>{\columncolor[HTML]{FFFFFF}}c 
>{\columncolor[HTML]{FFFFFF}}c 
>{\columncolor[HTML]{FFFFFF}}c 
>{\columncolor[HTML]{FFFFFF}}c |}
\hline
\textbf{Reorder}                   & \multicolumn{8}{c|}{\cellcolor[HTML]{FFFFFF}\textbf{INS-W}}                                                                                                                                                                                                                                                                                                                                                                                      & \multicolumn{4}{c|}{\cellcolor[HTML]{FFFFFF}\textbf{INS-D}}                                                                                                                                      \\ \hline
\textbf{Strategies/Dataset}                     & \multicolumn{2}{c|}{\cellcolor[HTML]{FFFFFF}\textbf{W\_1}}                                                            & \multicolumn{2}{c|}{\cellcolor[HTML]{FFFFFF}\textbf{W\_2}}                                                            & \multicolumn{2}{c|}{\cellcolor[HTML]{FFFFFF}\textbf{W\_3}}                                                            & \multicolumn{2}{c|}{\cellcolor[HTML]{FFFFFF}\textbf{W\_4}}               & \multicolumn{2}{c|}{\cellcolor[HTML]{FFFFFF}\textbf{D\_1}}                                                            & \multicolumn{2}{c|}{\cellcolor[HTML]{FFFFFF}\textbf{D\_2}}               \\ \hline
\multicolumn{1}{|l|}{\cellcolor[HTML]{FFFFFF}} & \multicolumn{1}{c|}{\cellcolor[HTML]{FFFFFF}\textbf{Acc}} & \multicolumn{1}{c|}{\cellcolor[HTML]{FFFFFF}\textbf{AUC}} & \multicolumn{1}{c|}{\cellcolor[HTML]{FFFFFF}\textbf{Acc}} & \multicolumn{1}{c|}{\cellcolor[HTML]{FFFFFF}\textbf{AUC}} & \multicolumn{1}{c|}{\cellcolor[HTML]{FFFFFF}\textbf{Acc}} & \multicolumn{1}{c|}{\cellcolor[HTML]{FFFFFF}\textbf{AUC}} & \multicolumn{1}{c|}{\cellcolor[HTML]{FFFFFF}\textbf{Acc}} & \textbf{AUC} & \multicolumn{1}{c|}{\cellcolor[HTML]{FFFFFF}\textbf{Acc}} & \multicolumn{1}{c|}{\cellcolor[HTML]{FFFFFF}\textbf{AUC}} & \multicolumn{1}{c|}{\cellcolor[HTML]{FFFFFF}\textbf{Acc}} & \textbf{AUC} \\ \hline
\textbf{Median}                                & \multicolumn{1}{c|}{\cellcolor[HTML]{FFFFFF}0.706}        & \multicolumn{1}{c|}{\cellcolor[HTML]{FFFFFF}0.782}        & \multicolumn{1}{c|}{\cellcolor[HTML]{FFFFFF}0.708}        & \multicolumn{1}{c|}{\cellcolor[HTML]{FFFFFF}0.758}        & \multicolumn{1}{c|}{\cellcolor[HTML]{FFFFFF}0.702}        & \multicolumn{1}{c|}{\cellcolor[HTML]{FFFFFF}0.758}        & \multicolumn{1}{c|}{\cellcolor[HTML]{FFFFFF}0.762}        & 0.811        & \multicolumn{1}{c|}{\cellcolor[HTML]{FFFFFF}0.787}        & \multicolumn{1}{c|}{\cellcolor[HTML]{FFFFFF}0.894}        & \multicolumn{1}{c|}{\cellcolor[HTML]{FFFFFF}0.775}        & 0.849        \\ \hline
\textbf{Simple kNN}                           & \multicolumn{1}{c|}{\cellcolor[HTML]{FFFFFF}0.694}        & \multicolumn{1}{c|}{\cellcolor[HTML]{FFFFFF}0.775}        & \multicolumn{1}{c|}{\cellcolor[HTML]{FFFFFF}0.681}        & \multicolumn{1}{c|}{\cellcolor[HTML]{FFFFFF}0.750}        & \multicolumn{1}{c|}{\cellcolor[HTML]{FFFFFF}0.723}        & \multicolumn{1}{c|}{\cellcolor[HTML]{FFFFFF}0.770}        & \multicolumn{1}{c|}{\cellcolor[HTML]{FFFFFF}0.729}        & 0.799        & \multicolumn{1}{c|}{\cellcolor[HTML]{FFFFFF}0.743}        & \multicolumn{1}{c|}{\cellcolor[HTML]{FFFFFF}0.834}        & \multicolumn{1}{c|}{\cellcolor[HTML]{FFFFFF}0.773}        & 0.845        \\ \hline
\textbf{Bounded kNN}                           & \multicolumn{1}{c|}{\cellcolor[HTML]{FFFFFF}0.677}        & \multicolumn{1}{c|}{\cellcolor[HTML]{FFFFFF}0.754}        & \multicolumn{1}{c|}{\cellcolor[HTML]{FFFFFF}0.695}        & \multicolumn{1}{c|}{\cellcolor[HTML]{FFFFFF}0.736}        & \multicolumn{1}{c|}{\cellcolor[HTML]{FFFFFF}0.670}        & \multicolumn{1}{c|}{\cellcolor[HTML]{FFFFFF}0.701}        & \multicolumn{1}{c|}{\cellcolor[HTML]{FFFFFF}0.683}        & 0.773        & \multicolumn{1}{c|}{\cellcolor[HTML]{FFFFFF}0.789}        & \multicolumn{1}{c|}{\cellcolor[HTML]{FFFFFF}0.880}        & \multicolumn{1}{c|}{\cellcolor[HTML]{FFFFFF}0.695}        & 0.773        \\ \hline
\textbf{MICE}                                  & \multicolumn{1}{c|}{\cellcolor[HTML]{FFFFFF}0.742}        & \multicolumn{1}{c|}{\cellcolor[HTML]{FFFFFF}0.811}        & \multicolumn{1}{c|}{\cellcolor[HTML]{FFFFFF}0.699}        & \multicolumn{1}{c|}{\cellcolor[HTML]{FFFFFF}0.756}        & \multicolumn{1}{c|}{\cellcolor[HTML]{FFFFFF}0.676}        & \multicolumn{1}{c|}{\cellcolor[HTML]{FFFFFF}0.772}        & \multicolumn{1}{c|}{\cellcolor[HTML]{FFFFFF}0.762}        & 0.806        & \multicolumn{1}{c|}{\cellcolor[HTML]{FFFFFF}0.785}        & \multicolumn{1}{c|}{\cellcolor[HTML]{FFFFFF}0.866}        & \multicolumn{1}{c|}{\cellcolor[HTML]{FFFFFF}0.739}        & 0.800        \\ \hline
\textbf{Autoencoder-Median}                    & \multicolumn{1}{c|}{\cellcolor[HTML]{FFFFFF}\textbf{0.766}}        & \multicolumn{1}{c|}{\cellcolor[HTML]{FFFFFF}0.821}        & \multicolumn{1}{c|}{\cellcolor[HTML]{FFFFFF}0.745}        & \multicolumn{1}{c|}{\cellcolor[HTML]{FFFFFF}\textbf{0.812}}        & \multicolumn{1}{c|}{\cellcolor[HTML]{FFFFFF}\textbf{0.735}}        & \multicolumn{1}{c|}{\cellcolor[HTML]{FFFFFF}\textbf{0.792}}        & \multicolumn{1}{c|}{\cellcolor[HTML]{FFFFFF}\textbf{0.797}}        & \textbf{0.855}        & \multicolumn{1}{c|}{\cellcolor[HTML]{FFFFFF}0.774}        & \multicolumn{1}{c|}{\cellcolor[HTML]{FFFFFF}0.885}        & \multicolumn{1}{c|}{\cellcolor[HTML]{FFFFFF}\textbf{0.785}}        & 0.855        \\ \hline
\textbf{Autoencoder-kNN}                       & \multicolumn{1}{c|}{\cellcolor[HTML]{FFFFFF}0.760}        & \multicolumn{1}{c|}{\cellcolor[HTML]{FFFFFF}\textbf{0.827}}        & \multicolumn{1}{c|}{\cellcolor[HTML]{FFFFFF}\textbf{0.745}}        & \multicolumn{1}{c|}{\cellcolor[HTML]{FFFFFF}0.807}        & \multicolumn{1}{c|}{\cellcolor[HTML]{FFFFFF}0.728}        & \multicolumn{1}{c|}{\cellcolor[HTML]{FFFFFF}0.781}        & \multicolumn{1}{c|}{\cellcolor[HTML]{FFFFFF}0.766}        & 0.837        & \multicolumn{1}{c|}{\cellcolor[HTML]{FFFFFF}\textbf{0.866}}        & \multicolumn{1}{c|}{\cellcolor[HTML]{FFFFFF}\textbf{0.908}}        & \multicolumn{1}{c|}{\cellcolor[HTML]{FFFFFF}0.777}        & \textbf{0.866}        \\ \hline
\end{tabular}
\Description{This table contains a comparison of different imputation strategies with the Reorder algorithm for the "within-person" depression prediction task when only training data is used for imputation to avoid data leakage. It also has the data availability after imputation for the strategies. We see that Autoencoder-Median and Autoencoder-kNN perform best or comparable to best in most cases}
}
\tiny
\label{tab:reorder-data-leakage}
\end{table*}

\subsubsection{Real Time Depression Prediction}
\begin{figure}[h]
  \centering
  \includegraphics[width=0.8\linewidth]{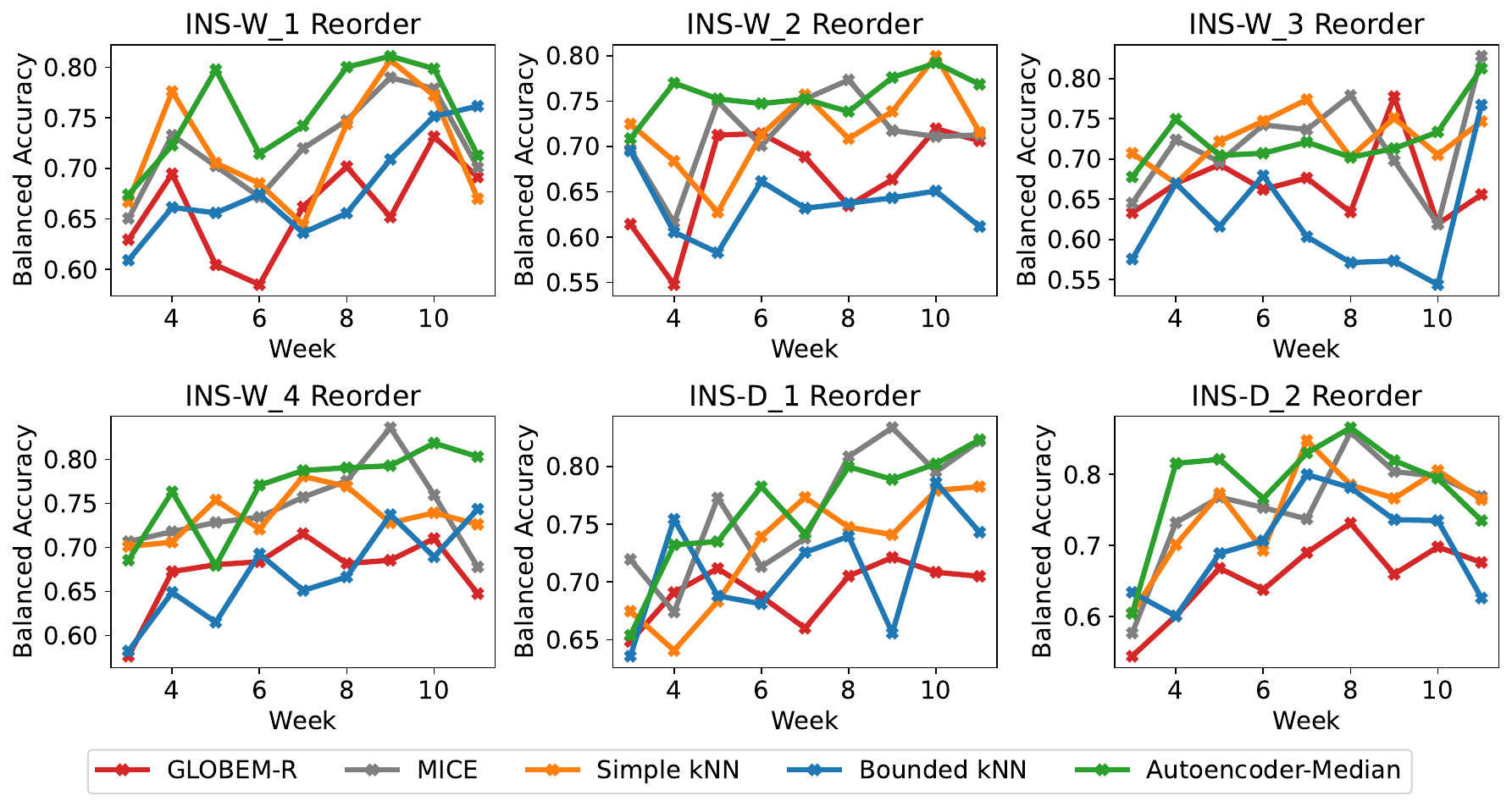}
  \caption{Balanced Accuracy comparison of various algorithms for real-time inductive imputation in Reorder algorithm.}
   \Description{Balanced Accuracy comparison of various algorithms for real-time inductive imputation in Reorder algorithm. Autoencoder Median performs better that all strategies for most cases for most of the weeks.}
  \label{fig:realtime_accuracies}
\end{figure}

To understand how these algorithms would perform in real-world systems that might require on-the-fly imputation for real-world predictions, we evaluated the various imputation strategies in a real-time simulations.
To evaluate the prediction objectives in a real-time scenario, we reframed our objective to predict depression at the end of each week in the study starting week 3. In Figure \ref{fig:realtime_accuracies}, we compared the balanced accuracies of the model trained with data imputed using different imputation accuracies. Despite slight fluctuations, we observed that the Autoencoder-Median consistently outperformed all other strategies for the majority of the weeks, suggesting it could be a reliable strategy even in real-time settings.



\label{subsec:inductive-realtime-imputation}

\section{Discussions}
\label{sec:discussions}






In this section, we discuss insights from our formative study involving thirteen researchers experienced in working with health and behavioral datasets. We tie together these insights to the findings from our case study on the GLOBEM datasets.

\subsection{Underreported \textit{and} Understudied}

Our formative interviews revealed that handling data missingness and imputation is not only under-reported in papers on passive health and behavioral sensing but is also understudied. Researchers often consider imputation a low-priority task compared to other aspects of study like developing sophisticated machine-learning models and other post-study analyses. They often use\textit{ ``ad-hoc''} and \textit{``hit and trial''} methods to choose imputation strategies, mostly relying on simple and sometimes on off-the-shelf strategies. This lack of a systematic evaluation of imputation strategies is due to time constraints and a greater focus on other aspects of the study. One way to encourage researchers to perform this systematic evaluation is to build a platform offering multiple imputation strategies where researchers can just plug in their datasets and explore different imputation strategies. As a first step, we make our implementations of imputation strategies publicly available on GitHub. In the long term, we can take inspiration from the GLOBEM Platform~\cite{xu2023globem}, where researchers can experiment with different depression detection algorithms and build a similar platform for imputation strategies. In our interviews, some researchers explicitly expressed this need. As P9 mentioned, \textit{``I always wanted to build a little tool where I can select the imputation method and quickly see the results, rather than to code things from scratch. With the limitation of time, the more complex implementation (of imputation strategies) gets deprioritized''}. This kind of platform can save researchers effort and time. As P12, who has mostly relied on simple imputation strategies, argued, \textit{``one of the ways to get me to use imputation strategies is to make it easy for me to use \ldots\ if somebody has a strategy and it's easy to use, and it's definitively shown to improve the model. Sure, I would use (it).''}

\subsection{Impact on Depression Prediction in GLOBEM Datasets} 
In our case study with the GLOBEM datasets, our proposed Autoencoder-based imputation strategy achieved up to 31\% improvements in AUROC for future depression prediction tasks. Although our Autoencoder-based strategy comes out promising and viable strategy, we do not claim it to be the best or the only imputation strategy for longitudinal health and behavior datasets as it warrants more investigation. As P11 noted, there is \textit{``no universal pattern''} with imputation strategies, and mentioned \textit{``It is very tricky to say one technique works better than others.''}. Our aim in the case study was to demonstrate that investing time in the development and evaluation of imputation algorithms can lead to performance improvements in the outcomes of the study (in this case, depression prediction). We hope our findings can encourage researchers to consider data imputation as a crucial part of the overall model development process and carefully explore multiple strategies to complete the datasets accurately.

\subsection{Reconstruction in GLOBEM Datasets}

Our evaluation of the reconstruction task revealed that based on the amputation strategies, there can be significant differences in the performance of imputation strategies. We saw different performances when we used MAR, MNAR (i), and MNAR (ii) strategies for amputing data. Replicating missingness through an amputation strategy, to be similar to the original observed data, is not feasible as multiple unobservable factors might interplay in causing missingness and is an inherent limitation with passive sensing data. Unlike the prediction tasks, where the study outcome is model performance on unseen data, the reconstruction task has no fixed outcome. Researchers seeking accurate reconstruction might have different goals, e.g., statistical modeling or post-hoc analyses. For instance, consider finding correlation between two features as the goal of the researcher. Choosing an imputation strategy that maximizes the correlation is an erroneous and misleading approach. P10 explained, \textit{``It (evaluating based on end outcomes) doesn't make sense for some kinds of analysis, for example, like if you are looking at correlations, and if you say that you are going to remove everything greater than three standard deviations and then you check correlation and you see it's not strong and then you go back to data and remove everything greater than two standard deviations. It doesn't make sense.''}

\subsection{Efficacy of Real-time Imputation}
While retrospective data analysis and model building post-study has been common in the past, the recent interest in personalized health and well-being predictions and interventions demands real-time and adaptive model building and predictions~\cite{mishra:detecting-receptivity, orzikulova2024time2stop}. We demonstrated a method to systematically evaluate the efficacy of imputation strategies in real-time settings, which can be adapted for future passive sensing studies. Traditional methods of data imputation often rely on static, retrospective approaches that may not adequately capture the dynamic nature of behavioral data in real-time. In contrast, inductive real-time data imputation leverages the longitudinal structure of the data, using past observations to impute data for subsequent points in time, e.g., the week after. By continuously updating imputation models as new data progressively becomes available, this approach could enable researchers designing future longitudinal studies to impute missing values and predict behavioral outcomes in near real-time, which could be helpful in cases where timely intervention is critical to the use-case of the study
\subsection{Common Pitfalls and Challenges with Imputation}

Through interviewing researchers and working on the case study with GLOBEM datasets, we synthesized a few common pitfalls and challenges with data imputation that researchers might encounter while working with longitudinal behavioral data.


\subsubsection{Deciding When to Impute}
As we demonstrated through our comparative analysis of multiple imputation strategies, data imputation plays a pivotal role in affecting the study outcomes. Despite this, there is no set standard on where the imputation step should fit in the data preprocessing pipeline for longitudinal datasets. In the past, researchers have applied imputation strategies after they divide the dataset into data batches/segments to train ML models, just using information present in those batches to impute. This is usually done as the necessary step without much consideration, as ML models do not accept missing values. While this delaying of imputation can work for datasets in other fields like Computer Vision, where the local context in the data batch is enough for estimating missing values, it does not work well for longitudinal behavioral datasets. Our results demonstrated this in the GLOBEM platform, where median imputation after dividing the dataset into data batches of 28 days (GLOBEM-R) performed way worse than median imputation on a participant’s entire dataset (Median). We advocate for integrating imputation at an early stage in the data pre-processing pipeline before subsetting the data into batches. This ensures that we use all available behavioral information to make precise estimates of missing values. Additionally, imputing values after dividing the dataset into input features can intensify the bias as we might miss out on the global behavioral patterns in the dataset.

\subsubsection{Data Leakage}
Researchers building prediction models often use the entirety of the dataset to complete the data and then divide the data into mutually exclusive train and test splits for model training and performance evaluation, respectively. The other way is that they first divide the data into train and test splits and then perform imputation. Data leakage occurs when some portion of test data flows into train data. While imputating on the entire dataset can be valid, it could cause data leakage, leading to an overestimation of the performance of models. P13 had a nuanced opinion,
\textit{``There are two camps on this, and I don't know which one I fall in. The first camp is you can only use the data from the validation set. You don't use the outcomes to impute because then you're going to get data leakage, and that makes total sense. But I think there is an argument to be made that when you're imputing you're trying to find the best possible like value for what that data point could have been, and so to get the best possible value, you use all of your data, including outcomes including, you know, all of the subjects.''}. While P11 preferred using data in the train set for imputation, explaining \textit{`` If we do some prediction tasks retrospectively and we take data from the future into account when doing the imputation, it's kind of like data leaking. So, in some past papers, I remember this has been done, which could lead to overestimated performance.''}. As in our case study, we strongly encourage researchers to do data leakage tests where they only impute using train data and keep the test data mutually exclusive to add robustness to their evaluations.

\subsubsection{Nature of Missingness}

Testing the nature of missingness in longitudinal behavioral datasets is challenging as there is no established and standard process. In our case study with GLOBEM datasets, we evaluated if the data is Missing Completely at Random (MCAR), using Little's MCAR Test~\cite{little1988test} on individual participants' data and found that for over 70\% of the participants, the missingness is not MCAR. Given the longitudinal nature of the data and potential within-participant variability, we thought it would be more appropriate to conduct individual MCAR tests. Contrary to our approach, some prior works on longitudinal passive sensing data conducted the MCAR test and reported that $p > 0.05$, thus concluding that the data was likely MCAR~\cite{nepal2024capturing, rashid2020predicting}. However, considering they only reported \textit{one} $p$-value, it appears they may have conducted one MCAR test for the entire dataset. 

The results from our approach indicate that data is not MCAR, which aligns with our expectations. Data missingness in longitudinal behavioral datasets may not be MCAR, as there are underlying reasons and behavioral patterns responsible for data gaps. This chain of thought was also supported in our interviews where researchers mentioned several behavioral and technical reasons for data missingness. For instance, P7 reported that participants would often take off sensors when they wanted to shower leading to data gaps. Understanding the nature of missingness is crucial as it may dictate the choice of imputation strategies. For instance, some commonly used imputation strategies like MICE have been shown to lead to biased estimates if data is not MAR \cite{goldberg2021data}.

.

\subsection{Limitations}

While our work highlights (1) the decision-making process and practices of researchers and (2) the importance of the development and evaluation of imputation strategies, It has some limitations:

\begin{itemize}

    
     \item \textit{Participants Recruitment}:
     For our interviews, we reached out to researchers working in the field of longitudinal health and behavioral passive-sensing, who submit in venues like CHI and Ubicomp. Insights from interviews might miss views and practices of other researchers working in adjacent fields, like behavioral sciences, psychology, and biomedical data sciences, and submitting to different venues. Moreover, it might also not cover the practices of researchers doing longitudinal studies in non-health-related fields.

    \item \textit{Data Missingness as a Behavioral Signal}:
    In our analyses, we treat missing sensor data as unavailability of behavioral signals, but, in some cases, missing data can be a strong behavioral signal; for instance, in substance use research, missing responses could be considered as positive cases of use events~\cite{mcpherson2012missing}. Some researchers in the past have modeled missing sensor data as a signal \cite{chikersal2021detecting}. As P13 explained \textit{`` We studied major depressive disorder; if a person missed a bunch of their surveys in a row, it could be because they lost their phone and it has nothing to do with their major depressive disorder, or it could be they're in a really depressed state, and they don't wanna answer surveys. And so there's the kind of notion of the missing data has signal in and of itself.''}

    \item {\textit{Generalizability}}: GLOBEM datasets have significant diversity in data dispersion across all datasets, which adds to the robustness of our evaluations. We, however, understand that the imputation strategies we developed and trends we observed might not generalize to other datasets. Moreover, GLOBEM datasets have data that are only available aggregated at the epoch level (morning, afternoon, evening, and night) and the day level, and thus, our results may not generalize other levels of data aggregations.

\end{itemize}





\section{Conclusion}
Missing data is inevitable in longitudinal passive sensing studies for health and behavior. Our formative interviews with researchers with experience in this domain revealed that most researchers assign imputation a low-priority step in their analysis and inference pipeline in favor of sophisticated model-building and post-study analyses. They often opt for using simple and off-the-shelf imputation strategies without comprehensively evaluating their impact on study outcomes. Furthermore, we found that processes and practices handling missing data and imputation strategies are often underreported in papers as well as understudied. Through a case study with publicly available depression detection datasets, we demonstrated ``Imputation Matters'' and highlight the importance of development and systematic evaluation of imputation strategies in significantly improving study outcomes.

%% file: ack.tex



\ifanonymized
 \relax
\else
 \section*{Acknowledgements}
 This research is
 supported by
the NIH National Institute of Drug Abuse under award number
  NIH/NIDA P30DA029926, 
 The views and conclusions contained herein are those of the authors and should not be interpreted as necessarily representing the official policies, either expressed or implied, of the sponsors.
 Any mention of specific companies or products does not imply any endorsement by the authors, by their employers, or by the sponsors.
\fi 

%% file: main.bbl

\begin{thebibliography}{74}


\ifx \showCODEN    \undefined \def \showCODEN     #1{\unskip}     \fi
\ifx \showDOI      \undefined \def \showDOI       #1{#1}\fi
\ifx \showISBNx    \undefined \def \showISBNx     #1{\unskip}     \fi
\ifx \showISBNxiii \undefined \def \showISBNxiii  #1{\unskip}     \fi
\ifx \showISSN     \undefined \def \showISSN      #1{\unskip}     \fi
\ifx \showLCCN     \undefined \def \showLCCN      #1{\unskip}     \fi
\ifx \shownote     \undefined \def \shownote      #1{#1}          \fi
\ifx \showarticletitle \undefined \def \showarticletitle #1{#1}   \fi
\ifx \showURL      \undefined \def \showURL       {\relax}        \fi
\providecommand\bibfield[2]{#2}
\providecommand\bibinfo[2]{#2}
\providecommand\natexlab[1]{#1}
\providecommand\showeprint[2][]{arXiv:#2}

\bibitem[Aravkin et~al\mbox{.}(2017)]%
        {aravkin2017generalized}
\bibfield{author}{\bibinfo{person}{Aleksandr Aravkin}, \bibinfo{person}{James~V Burke}, \bibinfo{person}{Lennart Ljung}, \bibinfo{person}{Aurelie Lozano}, {and} \bibinfo{person}{Gianluigi Pillonetto}.} \bibinfo{year}{2017}\natexlab{}.
\newblock \showarticletitle{Generalized Kalman smoothing: Modeling and algorithms}.
\newblock \bibinfo{journal}{\emph{Automatica}}  \bibinfo{volume}{86} (\bibinfo{year}{2017}), \bibinfo{pages}{63--86}.
\newblock


\bibitem[Assi et~al\mbox{.}(2023)]%
        {assi2023complex}
\bibfield{author}{\bibinfo{person}{Karim Assi}, \bibinfo{person}{Lakmal Meegahapola}, \bibinfo{person}{William Droz}, \bibinfo{person}{Peter Kun}, \bibinfo{person}{Amalia De~G{\"o}tzen}, \bibinfo{person}{Miriam Bidoglia}, \bibinfo{person}{Sally Stares}, \bibinfo{person}{George Gaskell}, \bibinfo{person}{Altangerel Chagnaa}, \bibinfo{person}{Amarsanaa Ganbold}, {et~al\mbox{.}}} \bibinfo{year}{2023}\natexlab{}.
\newblock \showarticletitle{Complex daily activities, country-level diversity, and smartphone sensing: A study in denmark, italy, mongolia, paraguay, and uk}. In \bibinfo{booktitle}{\emph{Proceedings of the 2023 CHI conference on human factors in computing systems}}. \bibinfo{pages}{1--23}.
\newblock


\bibitem[B{\"a}hr et~al\mbox{.}(2022)]%
        {bahr2022missing}
\bibfield{author}{\bibinfo{person}{Sebastian B{\"a}hr}, \bibinfo{person}{Georg-Christoph Haas}, \bibinfo{person}{Florian Keusch}, \bibinfo{person}{Frauke Kreuter}, {and} \bibinfo{person}{Mark Trappmann}.} \bibinfo{year}{2022}\natexlab{}.
\newblock \showarticletitle{Missing data and other measurement quality issues in mobile geolocation sensor data}.
\newblock \bibinfo{journal}{\emph{Social Science Computer Review}} \bibinfo{volume}{40}, \bibinfo{number}{1} (\bibinfo{year}{2022}), \bibinfo{pages}{212--235}.
\newblock


\bibitem[Barnett et~al\mbox{.}(2018)]%
        {barnett2018beyond}
\bibfield{author}{\bibinfo{person}{Ian Barnett}, \bibinfo{person}{John Torous}, \bibinfo{person}{Patrick Staples}, \bibinfo{person}{Matcheri Keshavan}, {and} \bibinfo{person}{Jukka-Pekka Onnela}.} \bibinfo{year}{2018}\natexlab{}.
\newblock \showarticletitle{Beyond smartphones and sensors: choosing appropriate statistical methods for the analysis of longitudinal data}.
\newblock \bibinfo{journal}{\emph{Journal of the American Medical Informatics Association}} \bibinfo{volume}{25}, \bibinfo{number}{12} (\bibinfo{year}{2018}), \bibinfo{pages}{1669--1674}.
\newblock


\bibitem[Bell and Fairclough(2014)]%
        {bell2014practical}
\bibfield{author}{\bibinfo{person}{Melanie~L Bell} {and} \bibinfo{person}{Diane~L Fairclough}.} \bibinfo{year}{2014}\natexlab{}.
\newblock \showarticletitle{Practical and statistical issues in missing data for longitudinal patient-reported outcomes}.
\newblock \bibinfo{journal}{\emph{Statistical methods in medical research}} \bibinfo{volume}{23}, \bibinfo{number}{5} (\bibinfo{year}{2014}), \bibinfo{pages}{440--459}.
\newblock


\bibitem[Bhattacharya et~al\mbox{.}(2024)]%
        {bhattacharya2024imputation}
\bibfield{author}{\bibinfo{person}{Sohini Bhattacharya}, \bibinfo{person}{Rahul Majethia}, \bibinfo{person}{Akshat Choube}, {and} \bibinfo{person}{Varun Mishra}.} \bibinfo{year}{2024}\natexlab{}.
\newblock \showarticletitle{Imputation Strategies for Longitudinal Behavioral Studies: Predicting Depression Using GLOBEM Datasets}. In \bibinfo{booktitle}{\emph{Companion of the 2024 on ACM International Joint Conference on Pervasive and Ubiquitous Computing}}. \bibinfo{pages}{736--742}.
\newblock


\bibitem[Bonell et~al\mbox{.}(2019)]%
        {bonell2019effects}
\bibfield{author}{\bibinfo{person}{Chris Bonell}, \bibinfo{person}{Emma Beaumont}, \bibinfo{person}{Matthew Dodd}, \bibinfo{person}{Diana~Ruth Elbourne}, \bibinfo{person}{Leonardo Bevilacqua}, \bibinfo{person}{Anne Mathiot}, \bibinfo{person}{Jennifer McGowan}, \bibinfo{person}{Joanna Sturgess}, \bibinfo{person}{Emily Warren}, \bibinfo{person}{Russell~M Viner}, {et~al\mbox{.}}} \bibinfo{year}{2019}\natexlab{}.
\newblock \showarticletitle{Effects of school environments on student risk-behaviours: evidence from a longitudinal study of secondary schools in England}.
\newblock \bibinfo{journal}{\emph{J Epidemiol Community Health}} \bibinfo{volume}{73}, \bibinfo{number}{6} (\bibinfo{year}{2019}), \bibinfo{pages}{502--508}.
\newblock


\bibitem[Burns et~al\mbox{.}(2011)]%
        {burns2011harnessing}
\bibfield{author}{\bibinfo{person}{Michelle~Nicole Burns}, \bibinfo{person}{Mark Begale}, \bibinfo{person}{Jennifer Duffecy}, \bibinfo{person}{Darren Gergle}, \bibinfo{person}{Chris~J Karr}, \bibinfo{person}{Emily Giangrande}, {and} \bibinfo{person}{David~C Mohr}.} \bibinfo{year}{2011}\natexlab{}.
\newblock \showarticletitle{Harnessing context sensing to develop a mobile intervention for depression}.
\newblock \bibinfo{journal}{\emph{Journal of medical Internet research}} \bibinfo{volume}{13}, \bibinfo{number}{3} (\bibinfo{year}{2011}), \bibinfo{pages}{e1838}.
\newblock


\bibitem[Campbell et~al\mbox{.}(2023)]%
        {campbell2023patient}
\bibfield{author}{\bibinfo{person}{Cynthia~I Campbell}, \bibinfo{person}{Ching-Hua Chen}, \bibinfo{person}{Sara~R Adams}, \bibinfo{person}{Asma Asyyed}, \bibinfo{person}{Ninad~R Athale}, \bibinfo{person}{Monique~B Does}, \bibinfo{person}{Saeed Hassanpour}, \bibinfo{person}{Emily Hichborn}, \bibinfo{person}{Melanie Jackson-Morris}, \bibinfo{person}{Nicholas~C Jacobson}, {et~al\mbox{.}}} \bibinfo{year}{2023}\natexlab{}.
\newblock \showarticletitle{Patient Engagement in a Multimodal Digital Phenotyping Study of Opioid Use Disorder}.
\newblock \bibinfo{journal}{\emph{Journal of Medical Internet Research}}  \bibinfo{volume}{25} (\bibinfo{year}{2023}), \bibinfo{pages}{e45556}.
\newblock


\bibitem[Chikersal et~al\mbox{.}(2021)]%
        {chikersal2021detecting}
\bibfield{author}{\bibinfo{person}{Prerna Chikersal}, \bibinfo{person}{Afsaneh Doryab}, \bibinfo{person}{Michael Tumminia}, \bibinfo{person}{Daniella~K Villalba}, \bibinfo{person}{Janine~M Dutcher}, \bibinfo{person}{Xinwen Liu}, \bibinfo{person}{Sheldon Cohen}, \bibinfo{person}{Kasey~G Creswell}, \bibinfo{person}{Jennifer Mankoff}, \bibinfo{person}{J~David Creswell}, {et~al\mbox{.}}} \bibinfo{year}{2021}\natexlab{}.
\newblock \showarticletitle{Detecting depression and predicting its onset using longitudinal symptoms captured by passive sensing: a machine learning approach with robust feature selection}.
\newblock \bibinfo{journal}{\emph{ACM Transactions on Computer-Human Interaction (TOCHI)}} \bibinfo{volume}{28}, \bibinfo{number}{1} (\bibinfo{year}{2021}), \bibinfo{pages}{1--41}.
\newblock


\bibitem[Choube et~al\mbox{.}(2024)]%
        {choube2024sesame}
\bibfield{author}{\bibinfo{person}{Akshat Choube}, \bibinfo{person}{Vedant~Das Swain}, {and} \bibinfo{person}{Varun Mishra}.} \bibinfo{year}{2024}\natexlab{}.
\newblock \showarticletitle{SeSaMe: A Framework to Simulate Self-Reported Ground Truth for Mental Health Sensing Studies}.
\newblock \bibinfo{journal}{\emph{arXiv preprint arXiv:2403.17219}} (\bibinfo{year}{2024}).
\newblock


\bibitem[Chow et~al\mbox{.}(2017)]%
        {chow2017using}
\bibfield{author}{\bibinfo{person}{Philip~I Chow}, \bibinfo{person}{Karl Fua}, \bibinfo{person}{Yu Huang}, \bibinfo{person}{Wesley Bonelli}, \bibinfo{person}{Haoyi Xiong}, \bibinfo{person}{Laura~E Barnes}, {and} \bibinfo{person}{Bethany~A Teachman}.} \bibinfo{year}{2017}\natexlab{}.
\newblock \showarticletitle{Using mobile sensing to test clinical models of depression, social anxiety, state affect, and social isolation among college students}.
\newblock \bibinfo{journal}{\emph{Journal of medical Internet research}} \bibinfo{volume}{19}, \bibinfo{number}{3} (\bibinfo{year}{2017}), \bibinfo{pages}{e62}.
\newblock


\bibitem[Christensen et~al\mbox{.}(2015)]%
        {christensen2015effect}
\bibfield{author}{\bibinfo{person}{Anne~I Christensen}, \bibinfo{person}{Ola Ekholm}, \bibinfo{person}{Peter~L Kristensen}, \bibinfo{person}{Finn~B Larsen}, \bibinfo{person}{Anker~L Vinding}, \bibinfo{person}{Charlotte Gl{\"u}mer}, {and} \bibinfo{person}{Knud Juel}.} \bibinfo{year}{2015}\natexlab{}.
\newblock \showarticletitle{The effect of multiple reminders on response patterns in a Danish health survey}.
\newblock \bibinfo{journal}{\emph{The European Journal of Public Health}} \bibinfo{volume}{25}, \bibinfo{number}{1} (\bibinfo{year}{2015}), \bibinfo{pages}{156--161}.
\newblock


\bibitem[Corbin and Strauss(1990)]%
        {corbin1990grounded}
\bibfield{author}{\bibinfo{person}{Juliet~M Corbin} {and} \bibinfo{person}{Anselm Strauss}.} \bibinfo{year}{1990}\natexlab{}.
\newblock \showarticletitle{Grounded theory research: Procedures, canons, and evaluative criteria}.
\newblock \bibinfo{journal}{\emph{Qualitative sociology}} \bibinfo{volume}{13}, \bibinfo{number}{1} (\bibinfo{year}{1990}), \bibinfo{pages}{3--21}.
\newblock


\bibitem[Cornet and Holden(2018)]%
        {cornet2018systematic}
\bibfield{author}{\bibinfo{person}{Victor~P Cornet} {and} \bibinfo{person}{Richard~J Holden}.} \bibinfo{year}{2018}\natexlab{}.
\newblock \showarticletitle{Systematic review of smartphone-based passive sensing for health and wellbeing}.
\newblock \bibinfo{journal}{\emph{Journal of biomedical informatics}}  \bibinfo{volume}{77} (\bibinfo{year}{2018}), \bibinfo{pages}{120--132}.
\newblock


\bibitem[Das~Swain et~al\mbox{.}(2022)]%
        {das2022semantic}
\bibfield{author}{\bibinfo{person}{Vedant Das~Swain}, \bibinfo{person}{Victor Chen}, \bibinfo{person}{Shrija Mishra}, \bibinfo{person}{Stephen~M Mattingly}, \bibinfo{person}{Gregory~D Abowd}, {and} \bibinfo{person}{Munmun De~Choudhury}.} \bibinfo{year}{2022}\natexlab{}.
\newblock \showarticletitle{Semantic Gap in Predicting Mental Wellbeing through Passive Sensing}. In \bibinfo{booktitle}{\emph{CHI Conference on Human Factors in Computing Systems}}. \bibinfo{pages}{1--16}.
\newblock


\bibitem[Das~Swain et~al\mbox{.}(2023)]%
        {das2020leveraging}
\bibfield{author}{\bibinfo{person}{Vedant Das~Swain}, \bibinfo{person}{Hyeokhyen Kwon}, \bibinfo{person}{Sonia Sargolzaei}, \bibinfo{person}{Bahador Saket}, \bibinfo{person}{Mehrab Bin~Morshed}, \bibinfo{person}{Kathy Tran}, \bibinfo{person}{Devashru Patel}, \bibinfo{person}{Yexin Tian}, \bibinfo{person}{Joshua Philipose}, \bibinfo{person}{Yulai Cui}, {et~al\mbox{.}}} \bibinfo{year}{2023}\natexlab{}.
\newblock \showarticletitle{Leveraging WiFi network logs to infer student collocation and its relationship with academic performance}.
\newblock \bibinfo{journal}{\emph{EPJ Data Science}} \bibinfo{volume}{12}, \bibinfo{number}{1} (\bibinfo{year}{2023}), \bibinfo{pages}{1--25}.
\newblock


\bibitem[{Das Swain} et~al\mbox{.}(2019)]%
        {DasSwain2019FitRoutine}
\bibfield{author}{\bibinfo{person}{Vedant {Das Swain}}, \bibinfo{person}{Manikanta~D. Reddy}, \bibinfo{person}{Kari~Anne Nies}, \bibinfo{person}{Louis Tay}, \bibinfo{person}{Munmun {De Choudhury}}, {and} \bibinfo{person}{Gregory~D. Abowd}.} \bibinfo{year}{2019}\natexlab{}.
\newblock \showarticletitle{{Birds of a Feather Clock Together: A Study of Person--Organization Fit Through Latent Activity Routines}}.
\newblock \bibinfo{journal}{\emph{Proc. ACM Hum.-Comput. Interact}} \bibinfo{number}{CSCW} (\bibinfo{year}{2019}).
\newblock


\bibitem[Das~Swain et~al\mbox{.}(2019)]%
        {dasswain2019multisensor}
\bibfield{author}{\bibinfo{person}{Vedant Das~Swain}, \bibinfo{person}{Koustuv Saha}, \bibinfo{person}{Hemang Rajvanshy}, \bibinfo{person}{Anusha Sirigiri}, \bibinfo{person}{Julie~M. Gregg}, \bibinfo{person}{Suwen Lin}, \bibinfo{person}{Gonzalo~J. Martinez}, \bibinfo{person}{Stephen~M. Mattingly}, \bibinfo{person}{Shayan Mirjafari}, \bibinfo{person}{Raghu Mulukutla}, {and} \bibinfo{person}{et al.}} \bibinfo{year}{2019}\natexlab{}.
\newblock \showarticletitle{A Multisensor Person-Centered Approach to Understand the Role of Daily Activities in Job Performance with Organizational Personas}.
\newblock \bibinfo{journal}{\emph{Proc. ACM IMWUT}} (\bibinfo{year}{2019}).
\newblock


\bibitem[DaSilva et~al\mbox{.}(2019)]%
        {dasilva2019correlates}
\bibfield{author}{\bibinfo{person}{Alex~W DaSilva}, \bibinfo{person}{Jeremy~F Huckins}, \bibinfo{person}{Rui Wang}, \bibinfo{person}{Weichen Wang}, \bibinfo{person}{Dylan~D Wagner}, {and} \bibinfo{person}{Andrew~T Campbell}.} \bibinfo{year}{2019}\natexlab{}.
\newblock \showarticletitle{Correlates of stress in the college environment uncovered by the application of penalized generalized estimating equations to mobile sensing data}.
\newblock \bibinfo{journal}{\emph{JMIR mHealth and uHealth}} \bibinfo{volume}{7}, \bibinfo{number}{3} (\bibinfo{year}{2019}), \bibinfo{pages}{e12084}.
\newblock


\bibitem[Fumagalli et~al\mbox{.}(2013)]%
        {fumagalli2013experiments}
\bibfield{author}{\bibinfo{person}{Laura Fumagalli}, \bibinfo{person}{Heather Laurie}, {and} \bibinfo{person}{Peter Lynn}.} \bibinfo{year}{2013}\natexlab{}.
\newblock \showarticletitle{Experiments with methods to reduce attrition in longitudinal surveys}.
\newblock \bibinfo{journal}{\emph{Journal of the Royal Statistical Society Series A: Statistics in Society}} \bibinfo{volume}{176}, \bibinfo{number}{2} (\bibinfo{year}{2013}), \bibinfo{pages}{499--519}.
\newblock


\bibitem[Gagnon-Audet et~al\mbox{.}(2022)]%
        {gagnon2022woods}
\bibfield{author}{\bibinfo{person}{Jean-Christophe Gagnon-Audet}, \bibinfo{person}{Kartik Ahuja}, \bibinfo{person}{Mohammad-Javad Darvishi-Bayazi}, \bibinfo{person}{Pooneh Mousavi}, \bibinfo{person}{Guillaume Dumas}, {and} \bibinfo{person}{Irina Rish}.} \bibinfo{year}{2022}\natexlab{}.
\newblock \showarticletitle{WOODS: Benchmarks for Out-of-Distribution Generalization in Time Series}.
\newblock \bibinfo{journal}{\emph{arXiv preprint arXiv:2203.09978}} (\bibinfo{year}{2022}).
\newblock


\bibitem[Goldberg et~al\mbox{.}(2021)]%
        {goldberg2021data}
\bibfield{author}{\bibinfo{person}{Simon~B Goldberg}, \bibinfo{person}{Daniel~M Bolt}, {and} \bibinfo{person}{Richard~J Davidson}.} \bibinfo{year}{2021}\natexlab{}.
\newblock \showarticletitle{Data missing not at random in mobile health research: Assessment of the problem and a case for sensitivity analyses}.
\newblock \bibinfo{journal}{\emph{Journal of Medical Internet Research}} \bibinfo{volume}{23}, \bibinfo{number}{6} (\bibinfo{year}{2021}), \bibinfo{pages}{e26749}.
\newblock


\bibitem[Griesler et~al\mbox{.}(2008)]%
        {griesler2008adolescents}
\bibfield{author}{\bibinfo{person}{Pamela~C Griesler}, \bibinfo{person}{Denise~B Kandel}, \bibinfo{person}{Christine Schaffran}, \bibinfo{person}{Mei-Chen Hu}, {and} \bibinfo{person}{Mark Davies}.} \bibinfo{year}{2008}\natexlab{}.
\newblock \showarticletitle{Adolescents' inconsistency in self-reported smoking: A comparison of reports in school and in household settings}.
\newblock \bibinfo{journal}{\emph{Public Opinion Quarterly}} \bibinfo{volume}{72}, \bibinfo{number}{2} (\bibinfo{year}{2008}), \bibinfo{pages}{260--290}.
\newblock


\bibitem[Gr{\"u}nerbl et~al\mbox{.}(2012)]%
        {grunerbl2012towards}
\bibfield{author}{\bibinfo{person}{Agnes Gr{\"u}nerbl}, \bibinfo{person}{Patricia Oleksy}, \bibinfo{person}{Gernot Bahle}, \bibinfo{person}{Christian Haring}, \bibinfo{person}{Jens Weppner}, {and} \bibinfo{person}{Paul Lukowicz}.} \bibinfo{year}{2012}\natexlab{}.
\newblock \showarticletitle{Towards smart phone based monitoring of bipolar disorder}. In \bibinfo{booktitle}{\emph{Proceedings of the Second ACM Workshop on Mobile Systems, Applications, and Services for HealthCare}}. \bibinfo{pages}{1--6}.
\newblock


\bibitem[Hern{\'a}ndez et~al\mbox{.}(2017)]%
        {hernandez2017data}
\bibfield{author}{\bibinfo{person}{Netzahualc{\'o}yotl Hern{\'a}ndez}, \bibinfo{person}{Luis~A Castro}, \bibinfo{person}{Jes{\'u}s Favela}, \bibinfo{person}{Layla Mich{\'a}n}, {and} \bibinfo{person}{Bert Arnrich}.} \bibinfo{year}{2017}\natexlab{}.
\newblock \showarticletitle{Data quality in mobile sensing datasets for pervasive healthcare}.
\newblock \bibinfo{journal}{\emph{Handbook of Large-Scale Distributed Computing in Smart Healthcare}} (\bibinfo{year}{2017}), \bibinfo{pages}{217--238}.
\newblock


\bibitem[Hovsepian et~al\mbox{.}(2015)]%
        {hovsepian2015cstress}
\bibfield{author}{\bibinfo{person}{Karen Hovsepian}, \bibinfo{person}{Mustafa Al'Absi}, \bibinfo{person}{Emre Ertin}, \bibinfo{person}{Thomas Kamarck}, \bibinfo{person}{Motohiro Nakajima}, {and} \bibinfo{person}{Santosh Kumar}.} \bibinfo{year}{2015}\natexlab{}.
\newblock \showarticletitle{cStress: towards a gold standard for continuous stress assessment in the mobile environment}. In \bibinfo{booktitle}{\emph{Proceedings of the 2015 ACM international joint conference on pervasive and ubiquitous computing}}. \bibinfo{pages}{493--504}.
\newblock


\bibitem[Huang(2015)]%
        {huang2015academic}
\bibfield{author}{\bibinfo{person}{Chiungjung Huang}.} \bibinfo{year}{2015}\natexlab{}.
\newblock \showarticletitle{Academic achievement and subsequent depression: A meta-analysis of longitudinal studies}.
\newblock \bibinfo{journal}{\emph{Journal of Child and Family Studies}}  \bibinfo{volume}{24} (\bibinfo{year}{2015}), \bibinfo{pages}{434--442}.
\newblock


\bibitem[Huang et~al\mbox{.}(2015)]%
        {huang2015insufficient}
\bibfield{author}{\bibinfo{person}{Jason~L Huang}, \bibinfo{person}{Mengqiao Liu}, {and} \bibinfo{person}{Nathan~A Bowling}.} \bibinfo{year}{2015}\natexlab{}.
\newblock \showarticletitle{Insufficient effort responding: examining an insidious confound in survey data.}
\newblock \bibinfo{journal}{\emph{Journal of Applied Psychology}} \bibinfo{volume}{100}, \bibinfo{number}{3} (\bibinfo{year}{2015}), \bibinfo{pages}{828}.
\newblock


\bibitem[Huckins et~al\mbox{.}(2019)]%
        {huckins2019fusing}
\bibfield{author}{\bibinfo{person}{Jeremy~F Huckins}, \bibinfo{person}{Alex~W DaSilva}, \bibinfo{person}{Rui Wang}, \bibinfo{person}{Weichen Wang}, \bibinfo{person}{Elin~L Hedlund}, \bibinfo{person}{Eilis~I Murphy}, \bibinfo{person}{Richard~B Lopez}, \bibinfo{person}{Courtney Rogers}, \bibinfo{person}{Paul~E Holtzheimer}, \bibinfo{person}{William~M Kelley}, {et~al\mbox{.}}} \bibinfo{year}{2019}\natexlab{}.
\newblock \showarticletitle{Fusing mobile phone sensing and brain imaging to assess depression in college students}.
\newblock \bibinfo{journal}{\emph{Frontiers in neuroscience}}  \bibinfo{volume}{13} (\bibinfo{year}{2019}), \bibinfo{pages}{248}.
\newblock


\bibitem[Ibrahim and Molenberghs(2009)]%
        {ibrahim2009missing}
\bibfield{author}{\bibinfo{person}{Joseph~G Ibrahim} {and} \bibinfo{person}{Geert Molenberghs}.} \bibinfo{year}{2009}\natexlab{}.
\newblock \showarticletitle{Missing data methods in longitudinal studies: a review}.
\newblock \bibinfo{journal}{\emph{Test}} \bibinfo{volume}{18}, \bibinfo{number}{1} (\bibinfo{year}{2009}), \bibinfo{pages}{1--43}.
\newblock


\bibitem[Kazijevs and Samad(2023)]%
        {kazijevs2023deep}
\bibfield{author}{\bibinfo{person}{Maksims Kazijevs} {and} \bibinfo{person}{Manar~D Samad}.} \bibinfo{year}{2023}\natexlab{}.
\newblock \showarticletitle{Deep imputation of missing values in time series health data: A review with benchmarking}.
\newblock \bibinfo{journal}{\emph{Journal of biomedical informatics}} (\bibinfo{year}{2023}), \bibinfo{pages}{104440}.
\newblock


\bibitem[Kramer et~al\mbox{.}(2020)]%
        {kramer:step-goals}
\bibfield{author}{\bibinfo{person}{Jan-Niklas Kramer}, \bibinfo{person}{Florian Künzler}, \bibinfo{person}{Varun Mishra}, \bibinfo{person}{Shawna~N Smith}, \bibinfo{person}{David Kotz}, \bibinfo{person}{Urte Scholz}, \bibinfo{person}{Elgar Fleisch}, {and} \bibinfo{person}{Tobias Kowatsch}.} \bibinfo{year}{2020}\natexlab{}.
\newblock \showarticletitle{{Which Components of a Smartphone Walking App Help Users to Reach Personalized Step Goals? Results From an Optimization Trial}}.
\newblock \bibinfo{journal}{\emph{Annals of Behavioral Medicine}} \bibinfo{volume}{54}, \bibinfo{number}{7} (\bibinfo{date}{March} \bibinfo{year}{2020}), \bibinfo{pages}{518--528}.
\newblock
\showISSN{0883-6612}
\urldef\tempurl%
\url{https://doi.org/10.1093/abm/kaaa002}
\showDOI{\tempurl}


\bibitem[Kristman et~al\mbox{.}(2005)]%
        {kristman2005methods}
\bibfield{author}{\bibinfo{person}{Vicki~L Kristman}, \bibinfo{person}{Michael Manno}, {and} \bibinfo{person}{Pierre C{\^o}t{\'e}}.} \bibinfo{year}{2005}\natexlab{}.
\newblock \showarticletitle{Methods to account for attrition in longitudinal data: do they work? A simulation study}.
\newblock \bibinfo{journal}{\emph{European journal of epidemiology}}  \bibinfo{volume}{20} (\bibinfo{year}{2005}), \bibinfo{pages}{657--662}.
\newblock


\bibitem[K{\"u}nzler et~al\mbox{.}(2019)]%
        {kunzler2019exploring}
\bibfield{author}{\bibinfo{person}{Florian K{\"u}nzler}, \bibinfo{person}{Varun Mishra}, \bibinfo{person}{Jan-Niklas Kramer}, \bibinfo{person}{David Kotz}, \bibinfo{person}{Elgar Fleisch}, {and} \bibinfo{person}{Tobias Kowatsch}.} \bibinfo{year}{2019}\natexlab{}.
\newblock \showarticletitle{Exploring the state-of-receptivity for mHealth interventions}.
\newblock \bibinfo{journal}{\emph{Proceedings of the ACM on Interactive, Mobile, Wearable and Ubiquitous Technologies}} \bibinfo{volume}{3}, \bibinfo{number}{4} (\bibinfo{year}{2019}), \bibinfo{pages}{1--27}.
\newblock


\bibitem[Laird(1988)]%
        {laird1988missing}
\bibfield{author}{\bibinfo{person}{Nan~M Laird}.} \bibinfo{year}{1988}\natexlab{}.
\newblock \showarticletitle{Missing data in longitudinal studies}.
\newblock \bibinfo{journal}{\emph{Statistics in medicine}} \bibinfo{volume}{7}, \bibinfo{number}{1-2} (\bibinfo{year}{1988}), \bibinfo{pages}{305--315}.
\newblock


\bibitem[Leeuw(2005)]%
        {leeuw2005dropout}
\bibfield{author}{\bibinfo{person}{E~de Leeuw}.} \bibinfo{year}{2005}\natexlab{}.
\newblock \showarticletitle{Dropout in longitudinal studies: Strategies to limit the problem}.
\newblock  (\bibinfo{year}{2005}).
\newblock


\bibitem[Little(1988)]%
        {little1988test}
\bibfield{author}{\bibinfo{person}{Roderick~JA Little}.} \bibinfo{year}{1988}\natexlab{}.
\newblock \showarticletitle{A test of missing completely at random for multivariate data with missing values}.
\newblock \bibinfo{journal}{\emph{Journal of the American statistical Association}} \bibinfo{volume}{83}, \bibinfo{number}{404} (\bibinfo{year}{1988}), \bibinfo{pages}{1198--1202}.
\newblock


\bibitem[Little(1995)]%
        {little1995modeling}
\bibfield{author}{\bibinfo{person}{Roderick~JA Little}.} \bibinfo{year}{1995}\natexlab{}.
\newblock \showarticletitle{Modeling the drop-out mechanism in repeated-measures studies}.
\newblock \bibinfo{journal}{\emph{Journal of the american statistical association}} \bibinfo{volume}{90}, \bibinfo{number}{431} (\bibinfo{year}{1995}), \bibinfo{pages}{1112--1121}.
\newblock


\bibitem[Loxton et~al\mbox{.}(2019)]%
        {loxton2019longitudinal}
\bibfield{author}{\bibinfo{person}{Deborah Loxton}, \bibinfo{person}{Jennifer Powers}, \bibinfo{person}{Natalie Townsend}, \bibinfo{person}{Melissa~L Harris}, {and} \bibinfo{person}{Peta Forder}.} \bibinfo{year}{2019}\natexlab{}.
\newblock \showarticletitle{Longitudinal inconsistency in responses to survey items that ask women about intimate partner violence}.
\newblock \bibinfo{journal}{\emph{BMC medical research methodology}}  \bibinfo{volume}{19} (\bibinfo{year}{2019}), \bibinfo{pages}{1--8}.
\newblock


\bibitem[Marsh and Yeung(1997)]%
        {marsh1997causal}
\bibfield{author}{\bibinfo{person}{Herbert~W Marsh} {and} \bibinfo{person}{Alexander~Seeshing Yeung}.} \bibinfo{year}{1997}\natexlab{}.
\newblock \showarticletitle{Causal effects of academic self-concept on academic achievement: Structural equation models of longitudinal data.}
\newblock \bibinfo{journal}{\emph{Journal of educational psychology}} \bibinfo{volume}{89}, \bibinfo{number}{1} (\bibinfo{year}{1997}), \bibinfo{pages}{41}.
\newblock


\bibitem[Mazumder et~al\mbox{.}(2010)]%
        {mazumder2010spectral}
\bibfield{author}{\bibinfo{person}{Rahul Mazumder}, \bibinfo{person}{Trevor Hastie}, {and} \bibinfo{person}{Robert Tibshirani}.} \bibinfo{year}{2010}\natexlab{}.
\newblock \showarticletitle{Spectral regularization algorithms for learning large incomplete matrices}.
\newblock \bibinfo{journal}{\emph{The Journal of Machine Learning Research}}  \bibinfo{volume}{11} (\bibinfo{year}{2010}), \bibinfo{pages}{2287--2322}.
\newblock


\bibitem[McDonald et~al\mbox{.}(2017)]%
        {mcdonald2017implications}
\bibfield{author}{\bibinfo{person}{Bennett McDonald}, \bibinfo{person}{Regine Haardoerfer}, \bibinfo{person}{Michael Windle}, \bibinfo{person}{Michael Goodman}, \bibinfo{person}{Carla Berg}, {et~al\mbox{.}}} \bibinfo{year}{2017}\natexlab{}.
\newblock \showarticletitle{Implications of attrition in a longitudinal web-based survey: an examination of college students participating in a tobacco use study}.
\newblock \bibinfo{journal}{\emph{JMIR public health and surveillance}} \bibinfo{volume}{3}, \bibinfo{number}{4} (\bibinfo{year}{2017}), \bibinfo{pages}{e7424}.
\newblock


\bibitem[McPherson et~al\mbox{.}(2012)]%
        {mcpherson2012missing}
\bibfield{author}{\bibinfo{person}{Sterling McPherson}, \bibinfo{person}{Celestina Barbosa-Leiker}, \bibinfo{person}{G~Leonard Burns}, \bibinfo{person}{Donelle Howell}, {and} \bibinfo{person}{John Roll}.} \bibinfo{year}{2012}\natexlab{}.
\newblock \showarticletitle{Missing data in substance abuse treatment research: current methods and modern approaches.}
\newblock \bibinfo{journal}{\emph{Experimental and clinical psychopharmacology}} \bibinfo{volume}{20}, \bibinfo{number}{3} (\bibinfo{year}{2012}), \bibinfo{pages}{243}.
\newblock


\bibitem[Mirjafari et~al\mbox{.}(2019)]%
        {mirjafari2019differentiating}
\bibfield{author}{\bibinfo{person}{Shayan Mirjafari}, \bibinfo{person}{Kizito Masaba}, \bibinfo{person}{Ted Grover}, \bibinfo{person}{Weichen Wang}, \bibinfo{person}{Pino Audia}, \bibinfo{person}{Andrew~T Campbell}, \bibinfo{person}{Nitesh~V Chawla}, \bibinfo{person}{Vedant~Das Swain}, \bibinfo{person}{Munmun~De Choudhury}, \bibinfo{person}{Anind~K Dey}, {et~al\mbox{.}}} \bibinfo{year}{2019}\natexlab{}.
\newblock \showarticletitle{Differentiating Higher and Lower Job Performers in the Workplace Using Mobile Sensing}.
\newblock \bibinfo{journal}{\emph{Proceedings of the ACM on Interactive, Mobile, Wearable and Ubiquitous Technologies}} \bibinfo{volume}{3}, \bibinfo{number}{2} (\bibinfo{year}{2019}), \bibinfo{pages}{37}.
\newblock


\bibitem[Mishra et~al\mbox{.}(2021)]%
        {mishra:detecting-receptivity}
\bibfield{author}{\bibinfo{person}{Varun Mishra}, \bibinfo{person}{Florian K\"{u}nzler}, \bibinfo{person}{Jan-Niklas Kramer}, \bibinfo{person}{Elgar Fleisch}, \bibinfo{person}{Tobias Kowatsch}, {and} \bibinfo{person}{David Kotz}.} \bibinfo{year}{2021}\natexlab{}.
\newblock \showarticletitle{{Detecting Receptivity for mHealth Interventions in the Natural Environment}}.
\newblock \bibinfo{journal}{\emph{Proc. ACM Interact. Mob. Wearable Ubiquitous Technol. (IMWUT)}} \bibinfo{volume}{5}, \bibinfo{number}{2} (\bibinfo{date}{June} \bibinfo{year}{2021}), \bibinfo{pages}{1--24}.
\newblock
\showISSN{2474-9567}
\urldef\tempurl%
\url{https://doi.org/10.1145/3463492}
\showDOI{\tempurl}


\bibitem[Mohs et~al\mbox{.}(2000)]%
        {mohs2000longitudinal}
\bibfield{author}{\bibinfo{person}{Richard~C Mohs}, \bibinfo{person}{James Schmeidler}, {and} \bibinfo{person}{Mosen Aryan}.} \bibinfo{year}{2000}\natexlab{}.
\newblock \showarticletitle{Longitudinal studies of cognitive, functional and behavioural change in patients with Alzheimer's disease}.
\newblock \bibinfo{journal}{\emph{Statistics in Medicine}} \bibinfo{volume}{19}, \bibinfo{number}{11-12} (\bibinfo{year}{2000}), \bibinfo{pages}{1401--1409}.
\newblock


\bibitem[M{\"o}ller et~al\mbox{.}(2013)]%
        {moller2013investigating}
\bibfield{author}{\bibinfo{person}{Andreas M{\"o}ller}, \bibinfo{person}{Matthias Kranz}, \bibinfo{person}{Barbara Schmid}, \bibinfo{person}{Luis Roalter}, {and} \bibinfo{person}{Stefan Diewald}.} \bibinfo{year}{2013}\natexlab{}.
\newblock \showarticletitle{Investigating self-reporting behavior in long-term studies}. In \bibinfo{booktitle}{\emph{Proceedings of the SIGCHI conference on human factors in computing systems}}. \bibinfo{pages}{2931--2940}.
\newblock


\bibitem[Mork and Nilsen(2012)]%
        {mork2012sleep}
\bibfield{author}{\bibinfo{person}{Paul~J Mork} {and} \bibinfo{person}{Tom~IL Nilsen}.} \bibinfo{year}{2012}\natexlab{}.
\newblock \showarticletitle{Sleep problems and risk of fibromyalgia: longitudinal data on an adult female population in Norway}.
\newblock \bibinfo{journal}{\emph{Arthritis \& Rheumatism}} \bibinfo{volume}{64}, \bibinfo{number}{1} (\bibinfo{year}{2012}), \bibinfo{pages}{281--284}.
\newblock


\bibitem[Nepal et~al\mbox{.}(2024)]%
        {nepal2024capturing}
\bibfield{author}{\bibinfo{person}{Subigya Nepal}, \bibinfo{person}{Wenjun Liu}, \bibinfo{person}{Arvind Pillai}, \bibinfo{person}{Weichen Wang}, \bibinfo{person}{Vlado Vojdanovski}, \bibinfo{person}{Jeremy~F Huckins}, \bibinfo{person}{Courtney Rogers}, \bibinfo{person}{Meghan~L Meyer}, {and} \bibinfo{person}{Andrew~T Campbell}.} \bibinfo{year}{2024}\natexlab{}.
\newblock \showarticletitle{Capturing the College Experience: A Four-Year Mobile Sensing Study of Mental Health, Resilience and Behavior of College Students during the Pandemic}.
\newblock \bibinfo{journal}{\emph{Proceedings of the ACM on Interactive, Mobile, Wearable and Ubiquitous Technologies}} \bibinfo{volume}{8}, \bibinfo{number}{1} (\bibinfo{year}{2024}), \bibinfo{pages}{1--37}.
\newblock


\bibitem[Newman(2014)]%
        {newman2014missing}
\bibfield{author}{\bibinfo{person}{Daniel~A Newman}.} \bibinfo{year}{2014}\natexlab{}.
\newblock \showarticletitle{Missing data: Five practical guidelines}.
\newblock \bibinfo{journal}{\emph{Organizational Research Methods}} \bibinfo{volume}{17}, \bibinfo{number}{4} (\bibinfo{year}{2014}), \bibinfo{pages}{372--411}.
\newblock


\bibitem[Obuchi et~al\mbox{.}(2020)]%
        {obuchi2020predicting}
\bibfield{author}{\bibinfo{person}{Mikio Obuchi}, \bibinfo{person}{Jeremy~F Huckins}, \bibinfo{person}{Weichen Wang}, \bibinfo{person}{Alex Dasilva}, \bibinfo{person}{Courtney Rogers}, \bibinfo{person}{Eilis Murphy}, \bibinfo{person}{Elin Hedlund}, \bibinfo{person}{Paul Holtzheimer}, \bibinfo{person}{Shayan Mirjafari}, {and} \bibinfo{person}{Andrew Campbell}.} \bibinfo{year}{2020}\natexlab{}.
\newblock \showarticletitle{Predicting brain functional connectivity using mobile sensing}.
\newblock \bibinfo{journal}{\emph{Proceedings of the ACM on interactive, mobile, wearable and ubiquitous technologies}} \bibinfo{volume}{4}, \bibinfo{number}{1} (\bibinfo{year}{2020}), \bibinfo{pages}{1--22}.
\newblock


\bibitem[Orzikulova et~al\mbox{.}(2024)]%
        {orzikulova2024time2stop}
\bibfield{author}{\bibinfo{person}{Adiba Orzikulova}, \bibinfo{person}{Han Xiao}, \bibinfo{person}{Zhipeng Li}, \bibinfo{person}{Yukang Yan}, \bibinfo{person}{Yuntao Wang}, \bibinfo{person}{Yuanchun Shi}, \bibinfo{person}{Marzyeh Ghassemi}, \bibinfo{person}{Sung-Ju Lee}, \bibinfo{person}{Anind~K Dey}, \bibinfo{person}{Xuhai Xu}, {et~al\mbox{.}}} \bibinfo{year}{2024}\natexlab{}.
\newblock \showarticletitle{Time2Stop: Adaptive and Explainable Human-AI Loop for Smartphone Overuse Intervention}.
\newblock \bibinfo{journal}{\emph{arXiv preprint arXiv:2403.05584}} (\bibinfo{year}{2024}).
\newblock


\bibitem[Philipson et~al\mbox{.}(2008)]%
        {philipson2008comparative}
\bibfield{author}{\bibinfo{person}{Peter~M Philipson}, \bibinfo{person}{Weang~Kee Ho}, {and} \bibinfo{person}{Robin Henderson}.} \bibinfo{year}{2008}\natexlab{}.
\newblock \showarticletitle{Comparative review of methods for handling drop-out in longitudinal studies}.
\newblock \bibinfo{journal}{\emph{Statistics in medicine}} \bibinfo{volume}{27}, \bibinfo{number}{30} (\bibinfo{year}{2008}), \bibinfo{pages}{6276--6298}.
\newblock


\bibitem[Rashid et~al\mbox{.}(2020)]%
        {rashid2020predicting}
\bibfield{author}{\bibinfo{person}{Haroon Rashid}, \bibinfo{person}{Sanjana Mendu}, \bibinfo{person}{Katharine~E Daniel}, \bibinfo{person}{Miranda~L Beltzer}, \bibinfo{person}{Bethany~A Teachman}, \bibinfo{person}{Mehdi Boukhechba}, {and} \bibinfo{person}{Laura~E Barnes}.} \bibinfo{year}{2020}\natexlab{}.
\newblock \showarticletitle{Predicting subjective measures of social anxiety from sparsely collected mobile sensor data}.
\newblock \bibinfo{journal}{\emph{Proceedings of the ACM on Interactive, Mobile, Wearable and Ubiquitous Technologies}} \bibinfo{volume}{4}, \bibinfo{number}{3} (\bibinfo{year}{2020}), \bibinfo{pages}{1--24}.
\newblock


\bibitem[Reyna et~al\mbox{.}(2020)]%
        {reyna2020early}
\bibfield{author}{\bibinfo{person}{Matthew~A Reyna}, \bibinfo{person}{Christopher~S Josef}, \bibinfo{person}{Russell Jeter}, \bibinfo{person}{Supreeth~P Shashikumar}, \bibinfo{person}{M~Brandon Westover}, \bibinfo{person}{Shamim Nemati}, \bibinfo{person}{Gari~D Clifford}, {and} \bibinfo{person}{Ashish Sharma}.} \bibinfo{year}{2020}\natexlab{}.
\newblock \showarticletitle{Early prediction of sepsis from clinical data: the PhysioNet/Computing in Cardiology Challenge 2019}.
\newblock \bibinfo{journal}{\emph{Critical care medicine}} \bibinfo{volume}{48}, \bibinfo{number}{2} (\bibinfo{year}{2020}), \bibinfo{pages}{210--217}.
\newblock


\bibitem[Saha et~al\mbox{.}(2019)]%
        {saha2019imputing}
\bibfield{author}{\bibinfo{person}{Koustuv Saha}, \bibinfo{person}{Manikanta~D Reddy}, \bibinfo{person}{Vedant das Swain}, \bibinfo{person}{Julie~M Gregg}, \bibinfo{person}{Ted Grover}, \bibinfo{person}{Suwen Lin}, \bibinfo{person}{Gonzalo~J Martinez}, \bibinfo{person}{Stephen~M Mattingly}, \bibinfo{person}{Shayan Mirjafari}, \bibinfo{person}{Raghu Mulukutla}, {et~al\mbox{.}}} \bibinfo{year}{2019}\natexlab{}.
\newblock \showarticletitle{Imputing missing social media data stream in multisensor studies of human behavior}. In \bibinfo{booktitle}{\emph{2019 8th International Conference on Affective Computing and Intelligent Interaction (ACII)}}. IEEE, \bibinfo{pages}{178--184}.
\newblock


\bibitem[Salthouse(2014)]%
        {salthouse2014selectivity}
\bibfield{author}{\bibinfo{person}{Timothy~A Salthouse}.} \bibinfo{year}{2014}\natexlab{}.
\newblock \showarticletitle{Selectivity of attrition in longitudinal studies of cognitive functioning}.
\newblock \bibinfo{journal}{\emph{Journals of Gerontology Series B: Psychological Sciences and Social Sciences}} \bibinfo{volume}{69}, \bibinfo{number}{4} (\bibinfo{year}{2014}), \bibinfo{pages}{567--574}.
\newblock


\bibitem[Sano et~al\mbox{.}(2018)]%
        {sano2018identifying}
\bibfield{author}{\bibinfo{person}{Akane Sano}, \bibinfo{person}{Sara Taylor}, \bibinfo{person}{Andrew~W McHill}, \bibinfo{person}{Andrew~JK Phillips}, \bibinfo{person}{Laura~K Barger}, \bibinfo{person}{Elizabeth Klerman}, {and} \bibinfo{person}{Rosalind Picard}.} \bibinfo{year}{2018}\natexlab{}.
\newblock \showarticletitle{Identifying objective physiological markers and modifiable behaviors for self-reported stress and mental health status using wearable sensors and mobile phones: observational study}.
\newblock \bibinfo{journal}{\emph{Journal of medical Internet research}} \bibinfo{volume}{20}, \bibinfo{number}{6} (\bibinfo{year}{2018}), \bibinfo{pages}{e210}.
\newblock


\bibitem[Santani et~al\mbox{.}(2018)]%
        {santani2018drinksense}
\bibfield{author}{\bibinfo{person}{Darshan Santani}, \bibinfo{person}{Florian Labhart}, \bibinfo{person}{Sara Landolt}, \bibinfo{person}{Emmanuel Kuntsche}, \bibinfo{person}{Daniel Gatica-Perez}, {et~al\mbox{.}}} \bibinfo{year}{2018}\natexlab{}.
\newblock \showarticletitle{DrinkSense: Characterizing youth drinking behavior using smartphones}.
\newblock \bibinfo{journal}{\emph{IEEE Transactions on Mobile Computing}} \bibinfo{volume}{17}, \bibinfo{number}{10} (\bibinfo{year}{2018}), \bibinfo{pages}{2279--2292}.
\newblock


\bibitem[Sarker et~al\mbox{.}(2016)]%
        {sarker2016finding}
\bibfield{author}{\bibinfo{person}{Hillol Sarker}, \bibinfo{person}{Matthew Tyburski}, \bibinfo{person}{Md~Mahbubur Rahman}, \bibinfo{person}{Karen Hovsepian}, \bibinfo{person}{Moushumi Sharmin}, \bibinfo{person}{David~H Epstein}, \bibinfo{person}{Kenzie~L Preston}, \bibinfo{person}{C~Debra Furr-Holden}, \bibinfo{person}{Adam Milam}, \bibinfo{person}{Inbal Nahum-Shani}, {et~al\mbox{.}}} \bibinfo{year}{2016}\natexlab{}.
\newblock \showarticletitle{Finding significant stress episodes in a discontinuous time series of rapidly varying mobile sensor data}. In \bibinfo{booktitle}{\emph{Proceedings of the 2016 CHI conference on human factors in computing systems}}. \bibinfo{pages}{4489--4501}.
\newblock


\bibitem[Teodorczuk et~al\mbox{.}(2007)]%
        {teodorczuk2007white}
\bibfield{author}{\bibinfo{person}{Andrew Teodorczuk}, \bibinfo{person}{John~T O'Brien}, \bibinfo{person}{Michael~J Firbank}, \bibinfo{person}{Leonardo Pantoni}, \bibinfo{person}{Anna Poggesi}, \bibinfo{person}{Timo Erkinjuntti}, \bibinfo{person}{Anders Wallin}, \bibinfo{person}{Lars-Olof Wahlund}, \bibinfo{person}{Alida Gouw}, \bibinfo{person}{Gunhild Waldemar}, {et~al\mbox{.}}} \bibinfo{year}{2007}\natexlab{}.
\newblock \showarticletitle{White matter changes and late-life depressive symptoms: longitudinal study}.
\newblock \bibinfo{journal}{\emph{The British Journal of Psychiatry}} \bibinfo{volume}{191}, \bibinfo{number}{3} (\bibinfo{year}{2007}), \bibinfo{pages}{212--217}.
\newblock


\bibitem[Triplet et~al\mbox{.}(2017)]%
        {triplet2017mail}
\bibfield{author}{\bibinfo{person}{Jacob~J Triplet}, \bibinfo{person}{Enesi Momoh}, \bibinfo{person}{Jennifer Kurowicki}, \bibinfo{person}{Leonardo~D Villarroel}, \bibinfo{person}{Tsun yee Law}, {and} \bibinfo{person}{Jonathan~C Levy}.} \bibinfo{year}{2017}\natexlab{}.
\newblock \showarticletitle{E-mail reminders improve completion rates of patient-reported outcome measures}.
\newblock \bibinfo{journal}{\emph{JSES Open Access}} \bibinfo{volume}{1}, \bibinfo{number}{1} (\bibinfo{year}{2017}), \bibinfo{pages}{25--28}.
\newblock


\bibitem[Troyanskaya et~al\mbox{.}(2001)]%
        {troyanskaya2001missing}
\bibfield{author}{\bibinfo{person}{Olga Troyanskaya}, \bibinfo{person}{Michael Cantor}, \bibinfo{person}{Gavin Sherlock}, \bibinfo{person}{Pat Brown}, \bibinfo{person}{Trevor Hastie}, \bibinfo{person}{Robert Tibshirani}, \bibinfo{person}{David Botstein}, {and} \bibinfo{person}{Russ~B Altman}.} \bibinfo{year}{2001}\natexlab{}.
\newblock \showarticletitle{Missing value estimation methods for DNA microarrays}.
\newblock \bibinfo{journal}{\emph{Bioinformatics}} \bibinfo{volume}{17}, \bibinfo{number}{6} (\bibinfo{year}{2001}), \bibinfo{pages}{520--525}.
\newblock


\bibitem[Van~Buuren and Groothuis-Oudshoorn(2011)]%
        {van2011mice}
\bibfield{author}{\bibinfo{person}{Stef Van~Buuren} {and} \bibinfo{person}{Karin Groothuis-Oudshoorn}.} \bibinfo{year}{2011}\natexlab{}.
\newblock \showarticletitle{mice: Multivariate imputation by chained equations in R}.
\newblock \bibinfo{journal}{\emph{Journal of statistical software}}  \bibinfo{volume}{45} (\bibinfo{year}{2011}), \bibinfo{pages}{1--67}.
\newblock


\bibitem[Vhaduri and Poellabauer(2017)]%
        {vhaduri2017design}
\bibfield{author}{\bibinfo{person}{Sudip Vhaduri} {and} \bibinfo{person}{Christian Poellabauer}.} \bibinfo{year}{2017}\natexlab{}.
\newblock \showarticletitle{Design factors of longitudinal smartphone-based health surveys}.
\newblock \bibinfo{journal}{\emph{Journal of Healthcare Informatics Research}}  \bibinfo{volume}{1} (\bibinfo{year}{2017}), \bibinfo{pages}{52--91}.
\newblock


\bibitem[Wang et~al\mbox{.}(2016)]%
        {wang2016crosscheck}
\bibfield{author}{\bibinfo{person}{Rui Wang}, \bibinfo{person}{Min~SH Aung}, \bibinfo{person}{Saeed Abdullah}, \bibinfo{person}{Rachel Brian}, \bibinfo{person}{Andrew~T Campbell}, \bibinfo{person}{Tanzeem Choudhury}, \bibinfo{person}{Marta Hauser}, \bibinfo{person}{John Kane}, \bibinfo{person}{Michael Merrill}, \bibinfo{person}{Emily~A Scherer}, {et~al\mbox{.}}} \bibinfo{year}{2016}\natexlab{}.
\newblock \showarticletitle{CrossCheck: toward passive sensing and detection of mental health changes in people with schizophrenia}. In \bibinfo{booktitle}{\emph{Proceedings of the 2016 ACM international joint conference on pervasive and ubiquitous computing}}. \bibinfo{pages}{886--897}.
\newblock


\bibitem[Wang et~al\mbox{.}(2014)]%
        {wang2014studentlife}
\bibfield{author}{\bibinfo{person}{Rui Wang}, \bibinfo{person}{Fanglin Chen}, \bibinfo{person}{Zhenyu Chen}, \bibinfo{person}{Tianxing Li}, \bibinfo{person}{Gabriella Harari}, \bibinfo{person}{Stefanie Tignor}, \bibinfo{person}{Xia Zhou}, \bibinfo{person}{Dror Ben-Zeev}, {and} \bibinfo{person}{Andrew~T Campbell}.} \bibinfo{year}{2014}\natexlab{}.
\newblock \showarticletitle{StudentLife: assessing mental health, academic performance and behavioral trends of college students using smartphones}. In \bibinfo{booktitle}{\emph{Proceedings of the 2014 ACM international joint conference on pervasive and ubiquitous computing}}. \bibinfo{pages}{3--14}.
\newblock


\bibitem[Wang et~al\mbox{.}(2018)]%
        {wang2018sensing}
\bibfield{author}{\bibinfo{person}{Weichen Wang}, \bibinfo{person}{Gabriella~M Harari}, \bibinfo{person}{Rui Wang}, \bibinfo{person}{Sandrine~R M{\"u}ller}, \bibinfo{person}{Shayan Mirjafari}, \bibinfo{person}{Kizito Masaba}, {and} \bibinfo{person}{Andrew~T Campbell}.} \bibinfo{year}{2018}\natexlab{}.
\newblock \showarticletitle{Sensing behavioral change over time: Using within-person variability features from mobile sensing to predict personality traits}.
\newblock \bibinfo{journal}{\emph{Proceedings of the ACM on Interactive, Mobile, Wearable and Ubiquitous Technologies}} \bibinfo{volume}{2}, \bibinfo{number}{3} (\bibinfo{year}{2018}), \bibinfo{pages}{1--21}.
\newblock


\bibitem[Wolke et~al\mbox{.}(2009)]%
        {wolke2009selective}
\bibfield{author}{\bibinfo{person}{Dieter Wolke}, \bibinfo{person}{Andrea Waylen}, \bibinfo{person}{Muthanna Samara}, \bibinfo{person}{Colin Steer}, \bibinfo{person}{Robert Goodman}, \bibinfo{person}{Tamsin Ford}, {and} \bibinfo{person}{Koen Lamberts}.} \bibinfo{year}{2009}\natexlab{}.
\newblock \showarticletitle{Selective drop-out in longitudinal studies and non-biased prediction of behaviour disorders}.
\newblock \bibinfo{journal}{\emph{The British Journal of Psychiatry}} \bibinfo{volume}{195}, \bibinfo{number}{3} (\bibinfo{year}{2009}), \bibinfo{pages}{249--256}.
\newblock


\bibitem[Xu et~al\mbox{.}(2023)]%
        {xu2023globem}
\bibfield{author}{\bibinfo{person}{Xuhai Xu}, \bibinfo{person}{Xin Liu}, \bibinfo{person}{Han Zhang}, \bibinfo{person}{Weichen Wang}, \bibinfo{person}{Subigya Nepal}, \bibinfo{person}{Yasaman Sefidgar}, \bibinfo{person}{Woosuk Seo}, \bibinfo{person}{Kevin~S Kuehn}, \bibinfo{person}{Jeremy~F Huckins}, \bibinfo{person}{Margaret~E Morris}, {et~al\mbox{.}}} \bibinfo{year}{2023}\natexlab{}.
\newblock \showarticletitle{GLOBEM: cross-dataset generalization of longitudinal human behavior modeling}.
\newblock \bibinfo{journal}{\emph{Proceedings of the ACM on Interactive, Mobile, Wearable and Ubiquitous Technologies}} \bibinfo{volume}{6}, \bibinfo{number}{4} (\bibinfo{year}{2023}), \bibinfo{pages}{1--34}.
\newblock


\bibitem[Xu et~al\mbox{.}(2022)]%
        {xu2022globem}
\bibfield{author}{\bibinfo{person}{Xuhai Xu}, \bibinfo{person}{Han Zhang}, \bibinfo{person}{Yasaman Sefidgar}, \bibinfo{person}{Yiyi Ren}, \bibinfo{person}{Xin Liu}, \bibinfo{person}{Woosuk Seo}, \bibinfo{person}{Jennifer Brown}, \bibinfo{person}{Kevin Kuehn}, \bibinfo{person}{Mike Merrill}, \bibinfo{person}{Paula Nurius}, {et~al\mbox{.}}} \bibinfo{year}{2022}\natexlab{}.
\newblock \showarticletitle{GLOBEM Dataset: Multi-Year Datasets for Longitudinal Human Behavior Modeling Generalization}.
\newblock \bibinfo{journal}{\emph{Advances in Neural Information Processing Systems}}  \bibinfo{volume}{35} (\bibinfo{year}{2022}), \bibinfo{pages}{24655--24692}.
\newblock


\bibitem[Young et~al\mbox{.}(2006)]%
        {young2006attrition}
\bibfield{author}{\bibinfo{person}{Anne~F Young}, \bibinfo{person}{Jennifer~R Powers}, {and} \bibinfo{person}{Sandra~L Bell}.} \bibinfo{year}{2006}\natexlab{}.
\newblock \showarticletitle{Attrition in longitudinal studies: who do you lose?}
\newblock \bibinfo{journal}{\emph{Australian and New Zealand journal of public health}} \bibinfo{volume}{30}, \bibinfo{number}{4} (\bibinfo{year}{2006}), \bibinfo{pages}{353--361}.
\newblock


\bibitem[Zhang et~al\mbox{.}(2021)]%
        {zhang2021relationship}
\bibfield{author}{\bibinfo{person}{Yuezhou Zhang}, \bibinfo{person}{Amos~A Folarin}, \bibinfo{person}{Shaoxiong Sun}, \bibinfo{person}{Nicholas Cummins}, \bibinfo{person}{Rebecca Bendayan}, \bibinfo{person}{Yatharth Ranjan}, \bibinfo{person}{Zulqarnain Rashid}, \bibinfo{person}{Pauline Conde}, \bibinfo{person}{Callum Stewart}, \bibinfo{person}{Petroula Laiou}, {et~al\mbox{.}}} \bibinfo{year}{2021}\natexlab{}.
\newblock \showarticletitle{Relationship between major depression symptom severity and sleep collected using a wristband wearable device: multicenter longitudinal observational study}.
\newblock \bibinfo{journal}{\emph{JMIR mHealth and uHealth}} \bibinfo{volume}{9}, \bibinfo{number}{4} (\bibinfo{year}{2021}), \bibinfo{pages}{e24604}.
\newblock


\end{thebibliography}
